\documentclass[%
reprint,
 amsmath,amssymb,
 aps,
floatfix,
]{revtex4-2}

\usepackage{graphicx}% Include figure files
\usepackage{dcolumn}% Align table columns on decimal point
\usepackage{bm}% bold math
\usepackage{float}
\usepackage{titlesec}

\titlespacing\section{0pt}{6pt plus 2pt minus 3pt}{6pt plus 2pt minus 3pt}
\usepackage{xcolor}

\usepackage{mleftright} %to get rid of the extra space around parentheses with \left and \right: write \mleft, \mright instead

\newcommand{\ket}[1]{\left|#1\right\rangle}
\newcommand{\bra}[1]{\left\langle#1\right|}
\newcommand{\be}{\begin{equation}}
\newcommand{\ee}{\end{equation}}
\newcommand{\bea}{\begin{eqnarray}}
\newcommand{\eea}{\end{eqnarray}}

\newcommand{\ketbra}[2]{\left| #1 \rangle \langle #2 \right|}

\newcommand{\tr}[1]{\text{Tr}\mleft( #1 \mright)}

\newcommand{\unit}[2]{#1\,#2}

\usepackage[colorlinks]{hyperref}% add hypertext capabilities
\hypersetup{%
	plainpages=true,
	breaklinks=true,% not default in dvips mode, so we must specify
	hypertexnames=false,%not ideal, but needed when pagenums duplicate (`i' vs. `1')
	pageanchor=true,
	colorlinks=true,
	linkcolor={blue},
	citecolor={red},
	urlcolor={blue},
	anchorcolor={black}
}

\newcommand{\figref}[1]{\mbox{Fig.~\ref{#1}}}
\newcommand{\figpanel}[2]{Fig.~\hyperref[#1]{\ref*{#1}#2}}
\newcommand{\figpanels}[3]{Fig.~\hyperref[#1]{\ref*{#1}#2,#3}}
\newcommand{\figpanelNoPrefix}[2]{\hyperref[#1]{\ref*{#1}#2}}
\renewcommand{\eqref}[1]{\mbox{Eq.~(\ref{#1})}}
\newcommand{\eqrefs}[2]{\mbox{Eqs.~(\ref{#1},\ref{#2})}}

\makeatletter
\def\maketitle{
\@author@finish
\title@column\titleblock@produce
\suppressfloats[t]}
\makeatother

\begin{document}

% \preprint{APS/123-QED}
\title{Extensive characterization of a family of efficient three-qubit gates at the coherence limit}
% \title{Generating families of quantum-entangled states via a native three-qubit gate}
% Force line breaks with \\

\author{Christopher W. Warren}
\email{warrenc@chalmers.se}
\author{Jorge Fern\'andez-Pend\'as}%
\author{Shahnawaz Ahmed}
\author{Tahereh Abad}
\author{Andreas Bengtsson}
\author{Janka Bizn\'arov\'a}
\author{Kamanasish Debnath}
\author{Xiu Gu}
\author{Christian Kri\v{z}an}
\author{Amr Osman}
\author{Anita Fadavi Roudsari}
\author{Per Delsing}
\author{G\"oran Johansson}
\author{Anton Frisk Kockum}
\author{Giovanna Tancredi}
\author{Jonas Bylander}
\affiliation{
 Department of Microtechnology and Nanoscience, Chalmers University of Technology\\
 412 96, Gothenburg, Sweden}

\date{\today}% It is always \today, today,

\begin{abstract}
  While all quantum algorithms can be expressed in terms of single-qubit and two-qubit gates, more expressive gate sets can help reduce the algorithmic depth. This is important in the presence of gate errors, especially those due to decoherence. Using superconducting qubits, we have implemented a three-qubit gate by simultaneously applying  two-qubit operations, thereby realizing a three-body interaction. This method straightforwardly extends to other quantum hardware architectures, requires only a ``firmware" upgrade to implement, and is faster than its constituent two-qubit gates. The three-qubit gate represents an entire family of operations, creating flexibility in quantum-circuit compilation. We demonstrate a gate fidelity of $97.90\%$, which is near the coherence limit of our device. We then generate two classes of entangled states, the GHZ and W states, by applying the new gate only once; in comparison, decompositions into the standard gate set would have a two-qubit gate depth of two and three, respectively. Finally, we combine characterization methods and analyze the experimental and statistical errors on the fidelity of the gates and of the target states. 
\end{abstract}
\maketitle

%\tableofcontents
% \section{Introduction}
Quantum algorithms are generally developed using single-qubit and two-qubit gates as the basis of the instruction set~\cite{Barenco1995,nielsen00}. All quantum algorithms can be decomposed into a minimal universal gate set consisting of such elements; however, this is not a requirement. Taking advantage of hardware-aware compilation or using larger-than-minimal gate sets can help reduce the algorithmic depth~\cite{Shi2020,Lacroix2020}: shallow quantum circuits are paramount in the presence of decoherence. Moreover, parameterized families of two-qubit interactions have enhanced the capabilities of quantum hardware by reducing the circuit depth~\cite{Lacroix2020}, improving the success probability of algorithms~\cite{Abrams2020}, or allowing more expressive gates tailored to specific problems~\cite{Arute2019, Foxen2020}. 

Three-qubit gates, such as the Toffoli and Fredkin gates, are central components of several quantum algorithms~\cite{PhysRevX.8.041015,PhysRevLett.121.010501,Martyn2021,Buhrman2001}. However, when only standard single- and two-qubit gate sets are available, compiling these three-qubit gates results in considerable overheads of additional gates~\cite{Barenco1995,Figgatt2019,Nengkun2013}. Having access to a native three-qubit gate implemented at the hardware level would therefore be beneficial, but unfortunately, the types of three-body interactions that naturally produce these gates can be difficult to engineer.
Toffoli gates have been implemented in a variety of implementations~\cite{PhysRevLett.102.040501,PhysRevLett.123.170503,Hendrickx2021,PhysRevApplied.14.014072}, but each had drawbacks that limited the scalability or the resulting fidelity of operation. Perhaps more importantly, the coherence limits of the operations were not discussed, making it difficult to determine whether the fidelities were intrinsic to the mode of operation or at the limit of what could be achieved by the particular implementation.

Recent work has modeled~\cite{XiuGu2021,Baker2022} or demonstrated~\cite{Kim2022} methods of implementing three-qubit gates through the simultaneous application of two-qubit gates. In particular, Gu et al.~\cite{XiuGu2021} analyzed a general model of three-body interactions generated by simultaneously driving two-qubit interactions through an intermediate state.  Such an implementation can be seen as a ``firmware" upgrade---meaning no changes to the underlying hardware, only to the control---and the physical gate set can be readily extended to include native three-qubit gates. 

In this work, we demonstrate the three-qubit Controlled-CPHASE-SWAP gate (CCZS), in a single application of the pulse controls, as proposed in Ref.~\cite{XiuGu2021}. It acts as a combination of Toffoli-like and Fredkin-like gates and implements an entire family of three-qubit interactions faster than the individual constituent operations. We provide a detailed analysis of the effect of decoherence and find that our gate fidelity is primarily coherence-limited. Notably, for specific parameters, the gate has the structure of a controlled fermionic SWAP gate, which we call a fermionic Fredkin (fFredkin), and which we use to demonstrate the rapid generation of entangled GHZ~\cite{Greenberger1989} and W~\cite{Dur2000} states in a single application of this multi-qubit operation.

The characterization of quantum processes and highly entangled states is non-trivial. Here we combine several methods used in the community (gate-set tomography, quantum process tomography, and quantum state tomography) to thoroughly analyze the achieved quantum-process and quantum-state fidelities as well as to trace the origins of the errors. Moreover, using sample bootstrapping, we determine the experimental and statistical errors attributed to these fidelities.
% \section{Experimental Methods}
\section*{Device Description}
\begin{figure*}[ht]
    \centering
    \includegraphics[width=0.95\linewidth]{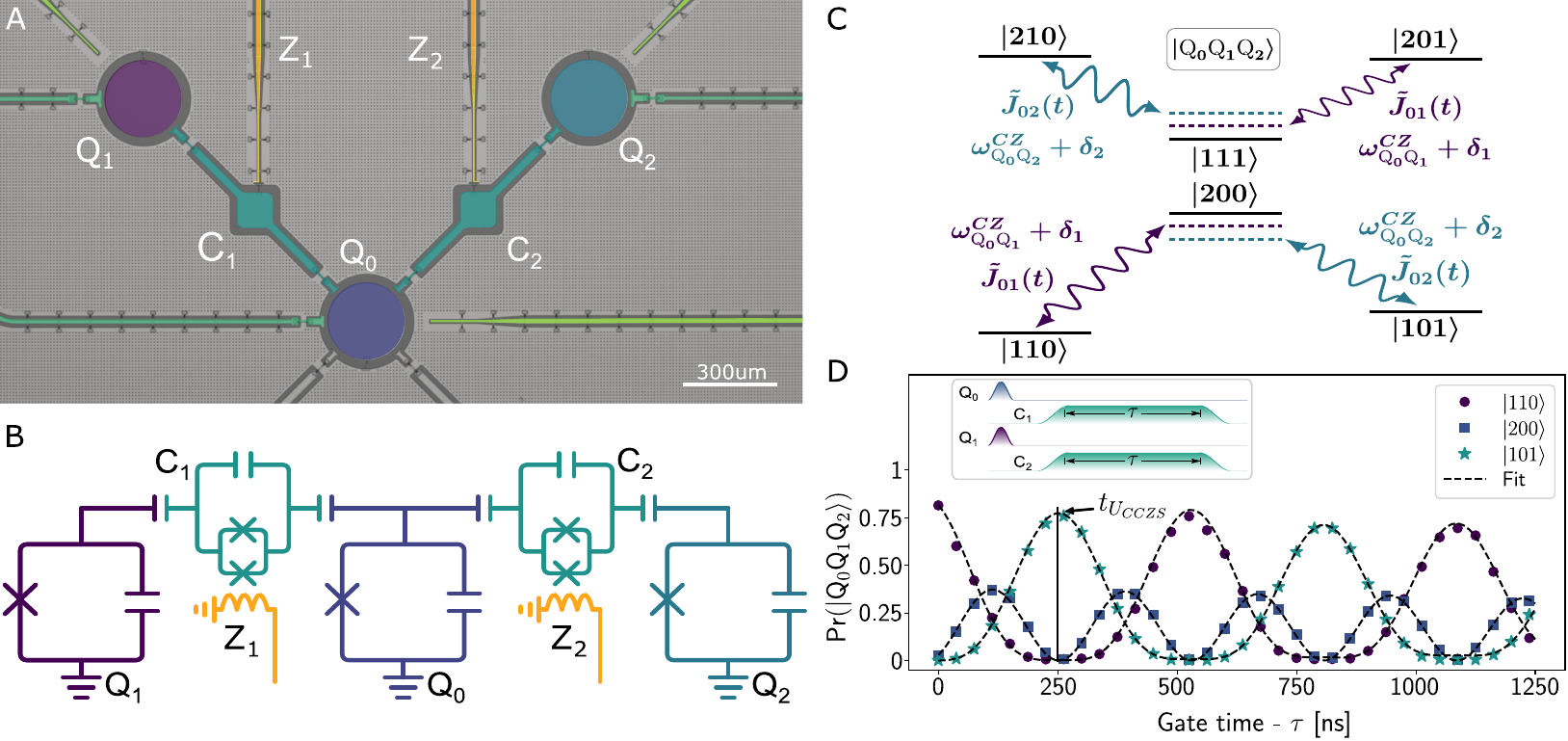}
    \caption{\textbf{(A)} Optical micrograph of the quantum processor. The three shown qubits ($Q_0$,$Q_1$,$Q_2$) are used in this work. The couplers ($C_1$,$C_2$) mediate coupling between neighboring qubits. \textbf{(B)} Reduced circuit diagram of the three-qubit device. \textbf{(C)} Energy levels of the $\Lambda$- and V-systems. Two AC pulses are applied at the CZ transition frequencies ($\omega^{CZ}_{Q_0 Q_1}$, $\omega^{CZ}_{Q_0 Q_2}$) on the couplers shared by $Q_0$, with driving strength $\tilde{J}_{0i}(t)$, activating an effective three-body interaction. The drives may be detuned from the true transition frequency due to miscalibration or Stark shifting during the drive.
    \textbf{(D)} Population transfer in the $\Lambda$-system after initializing $\ket{110}$. The $\Lambda$-system in (C) (bottom level structure) defines the SWAP component, whereas a round trip in the V-system (top level structure) causes the CCPHASE component (not shown).
    %This results in a $\sqrt{2}$ speed-up for a \unit{250}{ns} gate derived from constituent \unit{353}{ns} two-qubit gates.
    }
    \label{fig:Fig1}
\end{figure*}
Our experiment is conducted on three qubits ($Q_0, Q_1, Q_2$) of a five-qubit superconducting quantum processor as shown in \figpanel{fig:Fig1}{A}. The qubits are fixed-frequency transmon qubits~\cite{Koch2007}, each with individual control lines and readout resonators. Qubit-qubit interactions are mediated using flux-tunable transmon qubits $(C_1, C_2)$, referred to as couplers. The couplers are not considered in the computational space. This architecture with tunable couplers is flexible in that it allows for several types of two-qubit gates to be performed~\cite{Mckay2016,Sung2021,Ganzhorn2020}. The Hamiltonian of the circuit depicted in \figpanel{fig:Fig1}{B} is modeled as
\begin{eqnarray}
    \frac{H}{\hbar} &&= \sum_{i=0}^{2}\omega_i a_i^\dag a_i + \frac{\eta_i}{2}a_i^\dag a_i (a_i^\dag a_i - 1)\nonumber\\
      &&+ \sum_{j=1}^{2} \omega_{c_j}(\Phi_{j})b_j^\dag b_j + \frac{\eta_{c_j}}{2}b_j^\dag b_j (b_j^\dag b_j-1)\nonumber \\*
      &&+ \sum_{i,j} J_{ij}(a_i^\dag + a_i)(b_j^\dag + b_j).
\end{eqnarray}
The frequencies of the fixed-frequency qubits $i$ (couplers $j$) are given by $\omega_i$ ($\omega_{c_j}$ parameterized by magnetic flux $\Phi_j$). Each element of the system is modeled as a multi-level transmon with anharmonicity $\eta_i$ and annihilation (creation) operators $a_i$ ($a_i^\dagger$), $b_j$ ($b_j^\dagger$). Couplings between fixed-frequency qubits and couplers are denoted $J_{ij}$~\cite{supp}.

Interactions between pairs of qubits are generated by modulating the frequency of their shared tunable coupler~\cite{Mckay2016,Roth2017,Ganzhorn2020}. This is achieved by sending an AC signal to the SQUID of the coupler via a flux-bias line ($Z_1$ and $Z_2$ in \figpanel{fig:Fig1}{A}) of the form $\Phi_j(t) = \Phi_{b_j} + \Omega_j(t)\cos(\omega_{d_j} t + \phi_j)$, where $\Phi_{b_j}$ is the DC bias and $\Omega_j(t)$ is the pulse envelope. We use a cosine rise and fall of \unit{25}{ns} with flat time $\tau$. $\omega_{d_j}$ and $\phi_j$ are the AC driving frequency and phase of the signal, respectively. By modulating the coupler at the frequency difference between eigenstates of the qubits, as depicted in \figpanel{fig:Fig1}{C}, we selectively turn on interactions between pairs of qubits. In our case, we couple the $\ket{200}$ state of the central qubit to the $\ket{110}$ or $\ket{101}$ state. This interaction genereates a time-dependent effective coupling $\tilde J_{0i}(t)$ that implements a CPHASE gate in time $t_{g}^{CZ}=\pi/|\tilde J_{0i}|$, corresponding to a round trip from one of the initial computational states to $\ket{200}$ and back. 
\section*{Simultaneous Driving Dynamics}
Several proposals have been made for implementing effective three-qubit interactions~\cite{XiuGu2021,Kim2022,Baker2022}. In superconducting qubits, one such proposal has recently been demonstrated by applying simultaneous cross-resonance gates~\cite{Kim2022}. We follow an alternative schema based on the simultaneous parametric driving of tunable couplers as laid out in Ref.~\cite{XiuGu2021}. These results are general and can be readily applied to any doubly driven three-qubit system that has a similar level structure.

The simultaneous drives activate a $\Lambda$- and V-system within the two- and three-excitation manifolds (\figpanel{fig:Fig1}{C}). For these two systems, the dynamics are described by the same Hamiltonian
\begin{equation}
    \label{eqn:Eq2}
    H = \begin{bmatrix}
            -\delta_1 & \tilde{J}_{01} & 0  \\
            \tilde{J}_{01}^* & \delta_1 - \delta_2 & \tilde{J}_{02} \\
            0 & \tilde{J}_{02}^* & \delta_2
        \end{bmatrix},
\end{equation}
acting in the two subspaces $\{\ket{101}, \ket{200}, \ket{110}\}$ and $\{\ket{210}, \ket{111}, \ket{201}\}$.
%Again, we have the effective coupling rates $\tilde{J}_{0i}$ defined previously for the CZ transitions and
The terms $\delta_i$ represent the detuning of the respective drives from the true transition frequency. The simultaneous drives activate a CSWAP between $Q_1$ and $Q_2$ and additionally cause a CCPHASE, when $Q_0$ is in $\ket{1}$, in a time
\begin{equation}
    \label{eqn:Eq3}
    t_{g}^{CCZS} = \frac{\pi}{\sqrt{|\tilde J_{01}|^2 + |\tilde J_{02}|^2 + (\delta/2)^2}}.
\end{equation}
For convenience, we define $\delta_1=\delta_2=\delta$, but the dynamics of \eqref{eqn:Eq2} are solved generally to fit the dynamics of the $\Lambda$-system~\cite{supp}. The resulting three-qubit gate has the form
\begin{equation}
    U_{CCZS}(\theta, \phi, \gamma) = \ket{0}\bra{0}_0\otimes I_1 \otimes I_2 + \ket{1}\bra{1}_0\otimes U_{CZS}(\theta,\phi,\gamma)
\end{equation}
with
\begin{eqnarray}
    &&U_{CZS}(\theta, \phi, \gamma) =\\*
    && \begin{bmatrix}
            1 & 0 & 0 & 0 \\
            0 & -e^{i\gamma}\sin^2{\frac{\theta}{2}}+\cos^2{\frac{\theta}{2}} & e^{i(\frac{\gamma}{2}-\phi)}\cos{\frac{\gamma}{2}}\sin{\theta} & 0 \\
            0 & e^{i(\frac{\gamma}{2}+\phi)}\cos{\frac{\gamma}{2}}\sin{\theta}  & -e^{i\gamma}\cos^2{\frac{\theta}{2}}+\sin^2{\frac{\theta}{2}} & 0 \\
            0 & 0 & 0 & -e^{-i\gamma}
        \end{bmatrix}.\nonumber
\end{eqnarray}
The three-qubit gate has three parameters: the SWAP angle $\theta$; SWAP phase $\phi$; and CCPHASE phase $\gamma$, resulting in an entire family of three-qubit interactions. Experimentally, these angles are given by
\begin{eqnarray}
    \tan \frac{\theta}{2} = ~&&\frac{|\tilde J_{01}|}{|\tilde J_{02}|}e^{i\phi}, \\
    \gamma                = ~&&\frac{\pi\delta}{\sqrt{4(|\tilde J_{01}|^2 + |\tilde J_{02}|^2) + \delta^2}}.
\end{eqnarray}
The SWAP phase $\phi$ is controlled by the relative phase between the two AC flux drives, $\phi=\phi_1-\phi_2$. Virtual Z rotations are additionally applied to correct the frame of the qubits~\cite{Abrams2020,McKay2017} after the gate.

Of particular interest are the dynamics when the constituent two-qubit gates are of the same strength, i.e. $|\tilde J_{01}|=|\tilde J_{02}|$, and when the drives are on resonance with their corresponding transitions, $\delta=0$. With these parameters we obtain the gate
\begin{eqnarray}
    U_{CZS}(\pi/2, \phi, 0) = \begin{bmatrix}
                                1 & 0 & 0 & 0 \\
                                0 & 0 & e^{-i\phi} & 0 \\
                                0 & e^{i\phi}  & 0 & 0 \\
                                0 & 0 & 0 & -1
                            \end{bmatrix},\nonumber
\end{eqnarray}
which is implemented in a time $t_{g}^{CZ}/t_{g}^{CCZS}=\sqrt{2}$ times faster than the constituent two-qubit gates [\eqref{eqn:Eq3}]. For the instance of $\phi=0$ we obtain a controlled-fermionic-SWAP or fFredkin gate~\cite{Hashim2021}.
\begin{figure}
    \centering
    \includegraphics[width=0.95\linewidth]{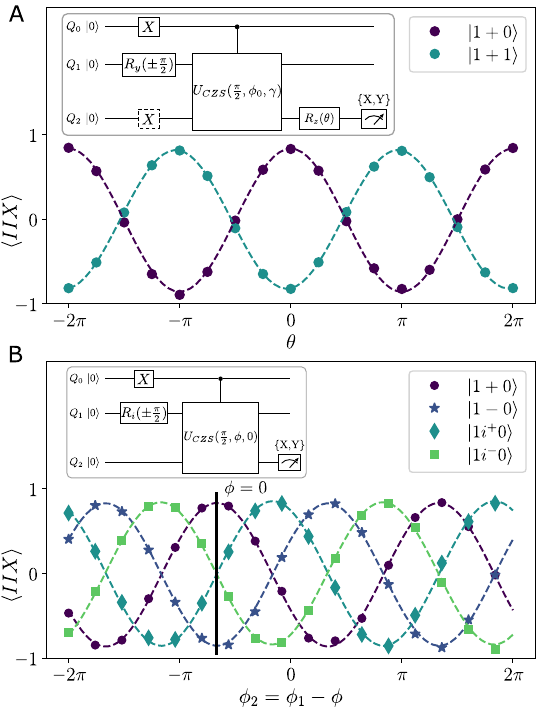}
    \caption{$\textbf{(A)}$ Calibration and verification of $\gamma$ (or equivalently $\delta$) similarly to typical CZ calibration~\cite{Ganzhorn2020}. We use $\ket{1+0}$ and $\ket{1+1}$ as probe states, as they are insensitive to the currently unknown SWAP phase, $\phi_0$, and oscillate $\gamma+\pi$ out of phase of one another. $\textbf{(B)}$ Performing a cross-Ramsey experiment using $Q_1$ prepared in $\ket{+}$ and use the three-qubit gate to SWAP the population to $Q_2$. We define $\phi=0$ from the phase that maximizes the expectation value in $I\otimes I\otimes X$. We additionally demonstrate full SWAP-phase control regardless of input state after calibration of $\gamma=0$.}
    \label{fig:Fig2}
\end{figure}
\section*{Determination of Gate Parameters}
We begin validating the driven model by first individually tuning up two pulses with equal effective coupling strengths $\tilde{J}_{0i}=\unit{2.833}{\text{MHz}}$ ($1/\tilde{J}_{0i}=$~\unit{353}{ns} pulse length) yielding $\theta=\pi/2$. We then apply the pulse sequence as depicted in the inset of \figpanel{fig:Fig1}{D}, preparing the state $\ket{110}$. 

In order to achieve the resonance condition, $\delta=0$, we perform two measurements in which we sweep the plateau of the simultaneous pulses as well as the frequency of one of the couplers' drives while keeping the other fixed. This produces oscillations in the $\Lambda$-system, which are fit to extract the frequency detunings $\delta_1$ and $\delta_2$~\cite{supp} as well as ensure equal coupling strengths. The population transfers of \figpanel{fig:Fig1}{D} correspond to a linecut in this 3D dataset (see \figref{fig:LandScape}) where $\delta_1=\delta_2=0$, with corresponding fit to the model in \eqref{eqn:Eq2}~\cite{supp}.

The resonance condition (and thus $\gamma=0$) is verified by performing the two experiments shown in the inset of \figpanel{fig:Fig2}{A}.
In the first, we prepare the state $\ket{1+0}$ and apply the $U_{CCZS}(\pi/2, \phi_0, \gamma)$ gate to swap the state for some, at the moment unknown, SWAP phase $\phi_0$. We then sweep the angle of a Z rotation on $Q_2$ and measure in either the X or Y basis. In the second experiment, we apply a NOT gate on the third qubit to prepare $\ket{1+1}$. The relative phase of the resulting superpositions are only sensitive to variations in $\gamma$. For any $\phi$, the two preparations oscillate $\pi$ out of phase when the resonance condition is achieved ($\delta_1=\delta_2=0$).

To determine the unknown SWAP phase $\phi_0$, we run the circuit in the inset of \figpanel{fig:Fig2}{B}. We prepare the input state $\ket{1+0}$ and then apply the simultaneous pulses while sweeping the phase of one of the AC drives relative to the other. The gate swaps the superposition state on $Q_1$ to $Q_2$, accumulating a phase at the difference between the individual phases of the two drives. We reference $\phi=0$ to the phase difference between drives which maximizes the expectation value $\langle IIX\rangle$ for the $\ket{1+0}$ state. Additionally, we demonstrate full control over $\phi$ by repeating the measurement with all eigenstates of X, $\ket{\pm}=\frac{1}{\sqrt{2}}(\ket{0}\pm\ket{1})$, and Y, $\ket{i^\pm}=\frac{1}{\sqrt{2}}(\ket{0}\pm i\ket{1})$, initialized on $Q_1$. We find coherent oscillations regardless of input state, demonstrating that we implement an entire family of three-qubit gates with the SWAP phase being a free parameter.

\section*{Gate Characterization}
\begin{figure*}[t]
    \centering
    \includegraphics[width=0.9\linewidth]{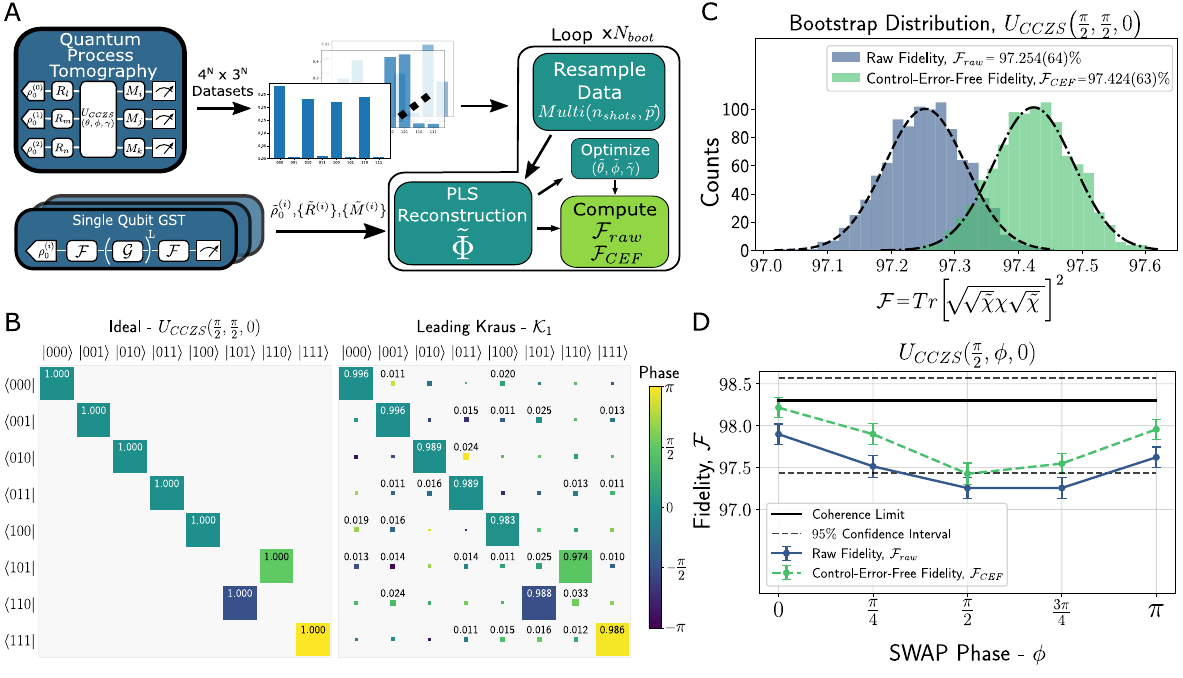}
    \caption{\textbf{(A)} Reconstruction procedure for QPT. Standard QPT is first performed to collect the $64 \times 27$ datasets comprising the reconstruction. Separately, single-qubit GST is performed to extract the noisy groundstates, rotation gates, and POVMs for the three qubits, which are used in the reconstruction to separate SPAM errors. \textbf{(B)} The ideal unitary of $U_{CCZS}(\frac{\pi}{2}, \frac{\pi}{2}, 0)$ and the leading Kraus (LK) matrix obtained from the Kraus operators. The LK matrix captures the majority of the dynamics of a noisy channel~\cite{CarignanDugas2019}. \textbf{(C)} Bootstrap distributions for a chosen SWAP angle of $\phi=\frac{\pi}{2}$ over $N_{boot}=1000$. The ``raw" process fidelity compares against the target unitary, whereas the control-error-free fidelity mitigates for imperfections in the $U_{CCZS}(\theta,\phi,\gamma)$ calibration. \textbf{(D)} Coherence limit of the three-qubit gate given the $T_1$, $T_2$ values with $95\%$ confidence interval~\cite{supp}, the raw fidelity of the reconstruction, and the control-error-free fidelity with $2\sigma$ error.}
    \label{fig:Fig3}
\end{figure*}
\begin{table*}[t]
    \begin{ruledtabular}
    %\color{red}
    \renewcommand{\arraystretch}{1.25}
    \begin{tabular}{cccccc}
        $(\theta,\phi,\gamma)$ & $\mathcal{F}_{raw}$ $[\%]$ & $\mathcal{F}_{CEF}$ $[\%]$ & $\tilde{\theta}/\pi$ & $\tilde{\phi}/\pi$ & $\tilde{\gamma}/\pi$ \\
        \hline
        $(\frac{\pi}{2},0,0)$              & $97.898(61)$ & $98.214(58)$ & $0.4969(9)$ & $-0.0160(7)$ & $-0.0374(9)$ \\
        $(\frac{\pi}{2},\frac{\pi}{4},0)$  & $97.514(65)$ & $97.898(63)$ & $0.4867(7)$ & $0.2378(7)$  & $-0.0411(9)$\\
        $(\frac{\pi}{2},\frac{\pi}{2},0)$  & $97.255(64)$ & $97.425(64)$ & $0.4954(8)$ & $0.4889(7)$ & $-0.0275(9)$\\
        $(\frac{\pi}{2},\frac{3\pi}{4},0)$ & $97.256(61)$ & $97.547(60)$ & $0.4928(7)$ & $0.7385(7)$ & $-0.0370(9)$\\
        $(\frac{\pi}{2},\pi,0)$            & $97.622(61)$ & $97.956(60)$ & $0.4988(7)$ & $0.9803(7)$ & $-0.0365(9)$\\
        \end{tabular}
        \end{ruledtabular}
        \caption{\label{tab:fid_table}Fidelity of the reconstruction for ideal target gate parameters $(\theta, \phi, \gamma)$ of the CCZS gate and the calibration-error-free fidelity. We report the angles best matching the reconstruction $(\tilde{\theta}, \tilde{\phi}, \tilde{\gamma})$ to understand how much of the fidelity is due to miscalibration. Errors in the angles due to miscalibration or drift deviate by $\le \frac{1}{75}\pi$, $\frac{1}{50}\pi$, and $\frac{1}{20}\pi$ for each of the parameters of the gate.}
\end{table*}

With the CCZS gate tuned up, we move on to characterization. We aim to obtain a measure of the fidelity of the gate independent of state-preparation-and-measurement (SPAM) errors. Several methods exist for this~\cite{Magesan2012,Erhard2019,Foxen2020}, but we seek more explicit information to trace whether the errors are the result of miscalibration, decoherence, or parasitic terms in the Hamiltonian. To achieve this we use quantum process tomography (QPT)~\cite{Chuang1997}. 

Process tomography is generally referenced to idealized state preparations, rotation operators and detectors, making it difficult to separate SPAM errors from the process being characterized~\cite{Merkel2013,Nielsen2021gatesettomography}. 
To remedy this, we modify the protocol by separately performing gate-set tomography (GST)~\cite{Nielsen2021gatesettomography}  on the single-qubit operations to obtain a model of the noisy initial states, noisy single-qubit rotations, and the single-qubit positive-operator valued measures (POVM) corresponding to readout. With these priors, we condition the QPT reconstruction to characterize our SPAM-free process~\cite{Geller2021}. The exact procedure is depicted in \figpanel{fig:Fig3}{A} and outlined in the supplement~\cite{supp}. 

For the reconstruction, we use the projected least squares (PLS) method~\cite{Knee2018,Surawystepney2021}, to obtain the Choi matrix representation~\cite{Jiang2013}, $\rho_\Phi$, of the quantum process, $\Phi$. The PLS method finds a least-squares estimate of the Choi matrix and then iteratively projects it into the space of completely positive trace-preserving (CPTP) maps which preserve physicality of the process.
% This algorithm iteratively projects to the space of completely positive trace-preserving (CPTP) maps representing physical processes.

In the Choi representation, a quantum channel evolves an input state $\rho$ as
\begin{equation}
    \label{eqn:eq8}
    \rho'=\Phi(\rho) = \mathrm{Tr}_a((\rho^T\otimes I_d)\rho_\Phi) ,
\end{equation}
where $I_d$ is the identity operator on a Hilbert space of dimension, $d$, equal to our system, and we take the partial trace over the input state's system.

From the reconstructed Choi matrix we can transform to any other representation of a quantum process such as to compute the process fidelity with the Chi matrix~\cite{korotkov2013error,UHLMANN1976273},
\begin{equation}
    \label{eqn:eq9}
    F(\chi, \tilde{\chi}) = \tr{\sqrt{\sqrt{\tilde \chi}\chi \sqrt{\tilde \chi}}}^2.
\end{equation}
$\chi$ and $\tilde{\chi}$ are Chi matrices representing two quantum processes.
%(we absorb the dimension of the superoperator to match the definition for density matrices). 
We can also obtain a quantum truth table, in that we obtain phase information as well as the classical mapping of input states by utilizing the Kraus representation to identify the dominant evolution~\cite{Wood2015,CarignanDugas2019} as shown in \figpanel{fig:Fig3}{B}.

However, the reconstruction only obtains a point estimate of the quantum process and the fidelity given our observations. Ideally, we would like to construct confidence intervals over the fidelity and over the space of possible reconstructions. We bootstrap the reconstruction by repeatedly sampling from the observed empirical distributions. Each newly sampled dataset represents possible experimental outcomes given the sample error of each QPT measurement~\cite{Efron1993,Home2009}. We report the average fidelity over the resulting distribution and the uncertainty, rather than the point estimate. The process fidelity of the three-qubit gate for all angles of $\phi$ is summarized in Table 1 for the \unit{250}{ns} three-qubit gate time that results from constituent \unit{353}{ns} two-qubit operations. We find that the process fidelity is near the coherence limit of $98.30\%$~\cite{supp}, as seen in \figpanel{fig:Fig3}{D}, and well within a $95\%$ confidence interval of the fluctuations of the coherence of our device.

There is an identifiable dependence of the fidelity on the SWAP phase $\phi$ as seen in \figpanel{fig:Fig3}{D} where the fidelity drops slightly until $\phi=\pi/2$ then increases again. While the decrease in fidelity lies within the typical fluctuations of the device, if we assume this trend is real, we can attempt to isolate the cause during the bootstrapping loop. We first perform the reconstruction and then optimize over the angles of an ideal CCZS gate to find the parameters $(\tilde{\theta}, \tilde{\phi}, \tilde{\gamma})$ that maximize the fidelity with respect to the reconstructed process (see~\figpanel{fig:Fig3}{C,D}). We term this fidelity the control-error-free fidelity, $\mathcal{F}_{CEF}$, and the fidelity from the reconstruction to the ideal target parameters as the ``raw" fidelity, $\mathcal{F}_{raw}$. If the $\phi$-dependence were purely due to drift in the controls or miscalibration, this optimization procedure should lift the dependency and flatten the fidelity along the coherence limit. 

While there are miscalibrations between the ideal target parameters (Table~\ref{tab:fid_table}), the seeming dependence on $\phi$ remains. Leakage to the couplers or population outside the computational space is exluded as leakage would be phase-indepedent. The sensitivity of exchange-like interactions to residual ZZ parasitic terms has been well documented in superconducting qubits~\cite{Ware2019,Noguchi2020,Xu2021,Ganzhorn2020,Sung2021} and could explain the resulting phase dependence. However, without further analysis it would be difficult to separate the influence of these parasitic terms from a fluctuation in coherence; such studies will be the subject of follow-up work.

\section*{Rapid Generation of Entangled States}
\begin{figure*}[ht]
    \centering
    \includegraphics[width=0.9\linewidth]{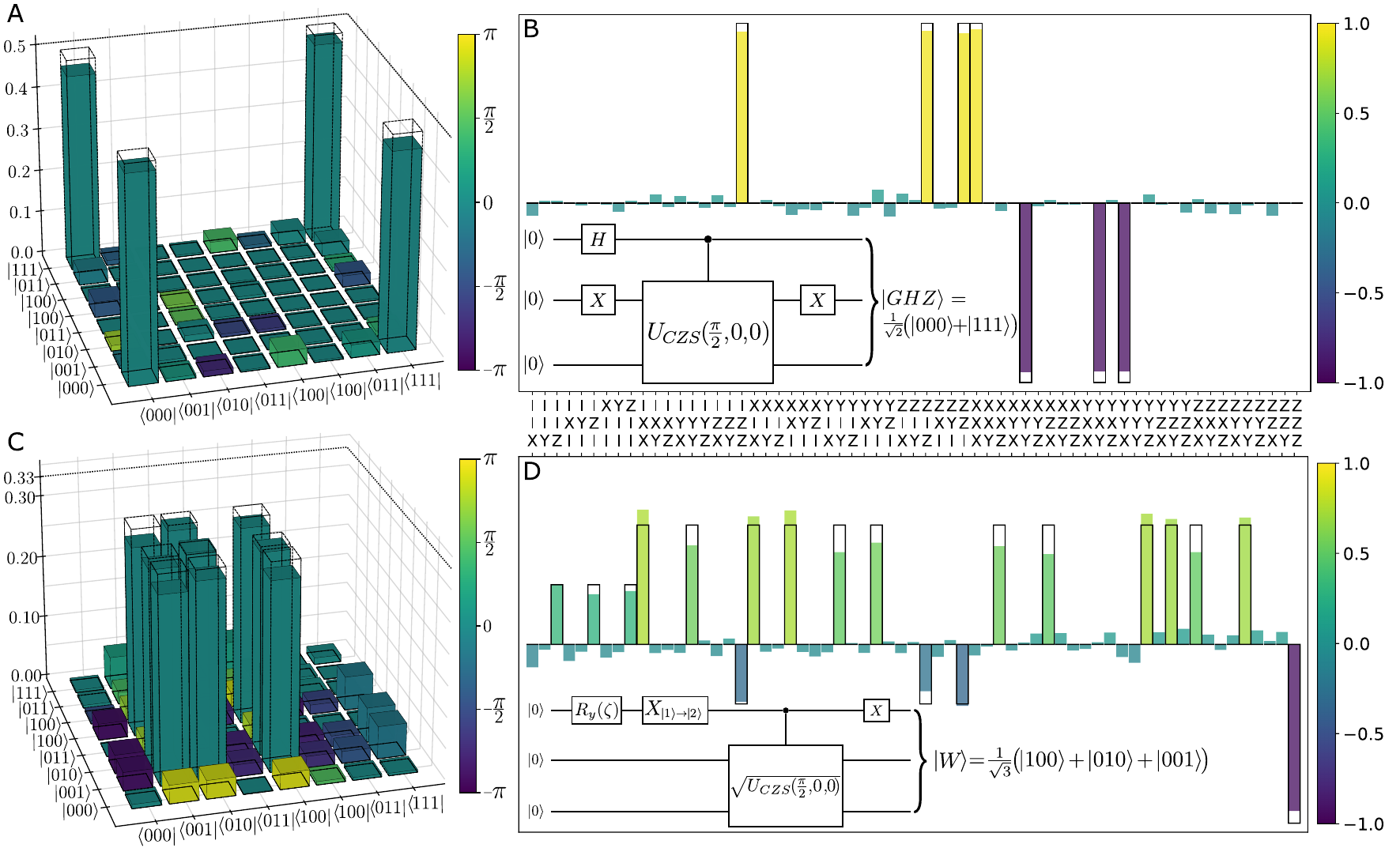}
    \caption{\textbf{(A} and \textbf{C)} Density matrix of the GHZ and W states with their magnitudes and phases plotted for each basis element. The theoretical values are plotted as the wireframes around the solid bars. \textbf{(B} and \textbf{D)} Expectation values of the different experimentally obtained Pauli observables and the ideal theoretical expectations and the corresponding circuits (insets) for generating the respective states.} 
     \label{fig:Fig4}
\end{figure*}
As a demonstration of the gate, we opt to use the CCZS gate in two different modes of operation. In the first, \figpanels{fig:Fig4}{A}{B}, we treat the three-qubit gate as acting solely within the computational subspace and prepare a GHZ state, $(\ket{000}+\ket{111})/\sqrt{2}$, in a single application of the CCZS. With only two-qubit operations, this requires the application of two sequential two-qubit gates. We achieve a state fidelity of $95.56(16)\%$ (using \eqref{eqn:eq9} for the fidelity) after mitigating measurement errors.

In the second case, \figpanels{fig:Fig4}{C}{D}, we allow for evolution outside of the computational subspace and apply the CCZS gate for approximately half the time (denoted a $\sqrt{\text{CCZS}}$ gate) using the same gate parameters. This alternative implementation leverages the qutrit space to rapidly generate the W state.

Hence, in our circuit (see inset of \figpanel{fig:Fig4}{D}) we first prepare the state $\sqrt{1/3}\ket{000}+\sqrt{2/3}\ket{100}$ by applying a rotation $R_{y}(2 \arccos{\sqrt{1/3}})$ to $Q_0$. We then apply a calibrated $X_{1\rightarrow 2}$ pulse to perform a NOT operation in the qutrit space and map the population in $\ket{100}$ to $\ket{200}$. From here the application of the $\sqrt{CCZS}$ gate divides the population in $\ket{200}$ to states $\ket{101}$ and $\ket{110}$, resulting in the state $(\ket{000}+e^{i\phi_1}\ket{110}+e^{i\phi_2}\ket{101})/\sqrt{3}$. A final NOT gate applied to $Q_0$ completes the generation of the W state up to locally correctable phases with single-qubit $R_z$ gates. This three-qutrit gate is implemented in \unit{133}{ns} and generates the W state with a fidelity of $94.71(21)\%$. The generation of this state using conventional two-qubit control would require a depth of three sequential two-qubit gates (see \figref{fig:circuitcomp}).
\section*{Discussion and Outlook}
We demonstrated a single-step implementation of a family of three-qubit gates based on simultaneous driving of transitions to an intermediate eigenstate. These gates combine aspects of Toffoli and Fredkin gates resulting in an operation that we denote controlled-CPHASE-SWAP, or CCZS. Our approach is extensible, as it can be implemented across larger qubit systems and other quantum-computing implementations. For this reason, the three-qubit gate represents a ``firmware" upgrade of existing systems: the only requirement is the simultaneous driving of transitions to a common eigenstate in a multi-qubit system. The calibration uses existing two-qubit gate strategies and can be straightforwardly applied to other systems. This results in process fidelities approaching the coherence limit of our device of $\sim 98\%$.

%We find that the theory of a simultaneously driven $\lambda$-system describes the operations of the three-qubit interaction and predicts a $\sqrt{2}$ speed-up when compared to the interactions generated by the individual two-qubit drives. Using this three-qubit gate we demonstrate full control over the SWAP phase.

Applying the CCZS gate to our hardware allowed us to rapidly prepare two different classes of entangled states. We therefore envision that this gate can be used to augment existing gate sets and leverage the multi-qubit nature to aid in compilation of quantum algorithms. In particular, the rapid generation of GHZ states would facilitate the rapid creation and distillation of larger entangled states~\cite{deBone2020}, which can then be used as resources to demonstrate the power of unbounded quantum fanout gates~\cite{v001a005} experimentally. The CCZS gate can also be used to generate more familiar three-qubit gates such as the iFredkin, with the addition of a single CZ gate~\cite{XiuGu2021}, or for $\phi=0$, a fermionic Fredkin gate. Beyond the computational basis, this gate can be used to augment gate sets in qutrit systems, which has been a comparatively unexplored field.

% We characterize the resulting gate using quantum process tomography (QPT) augmented by first performing gate set tomography (GST) and using these results to mitigate the state-preparation and measurement errors (SPAM) in the reconstruction. We find that the process fidelity is at the predicted coherence limit. However to verify the reconstruction we use the $U_{CCZS}$ and the $\sqrt{U_{CCZS}}$ to rapidly generate the multi-qubit entangled GHZ and W states respectively. We find that in \textcolor{red}{200ns} we can generate a GHZ state with a fidelity of \textcolor{red}{$xxx\%$} and a fidelity for the W state of \textcolor{red}{$xxx\%$} in as little as \textcolor{red}{133ns}.

% \textcolor{red}{Need a strong finisher here. Ideally some ideas on how this could be used in some algorithm or arguments about how this gate can be used to reduced the depth of a circuit in compilation. Maybe just cite Xiu's paper and discuss how this could be applied practically?}

\begin{acknowledgments}
We are grateful to the Quantum Device Lab at ETH Z\"urich for sharing their designs of sample holder and printed circuit board. We would also like to thank Morten Kjaergaard, Oliver Hahn, Kenneth Rudinger, Stefan Seritan, Corey Ostrove, Christian Andersen, and Derek Jouppi for insightful discussions. Samples were fabricated at Myfab Chalmers. We acknowledge financial support from the Knut and Alice Wallenberg Foundation through the Wallenberg Center for Quantum Technology (WACQT), the Swedish Research Council, and the EU Flagship on Quantum Technology H2020-FETFLAG-2018-03 project 820363 OpenSuperQ.
\end{acknowledgments}
\newpage
\bibliography{CCZS_3QGate}% Produces the bibliography via BibTeX.

\renewcommand{\thesection}{SUPPLEMENTARY NOTE \arabic{section}}
\renewcommand{\thefigure}{S\arabic{figure}}
\renewcommand{\thetable}{S\arabic{table}}
\renewcommand{\theequation}{S\arabic{equation}}
\setcounter{figure}{0}
\setcounter{table}{0}
\setcounter{equation}{0}

% \appendix
\clearpage
\title{Supplementary Information: Extensive characterization of a family of efficient three-qubit gates at the coherence limit}
\maketitle
\onecolumngrid
\titlespacing\section{0pt}{16pt plus 4pt minus 4pt}{16pt plus 4pt minus 4pt}
% \section{Supplementary Material}
\section{Measurement setup}
\begin{figure}[t]
    \centering
    \includegraphics[width=0.95\linewidth]{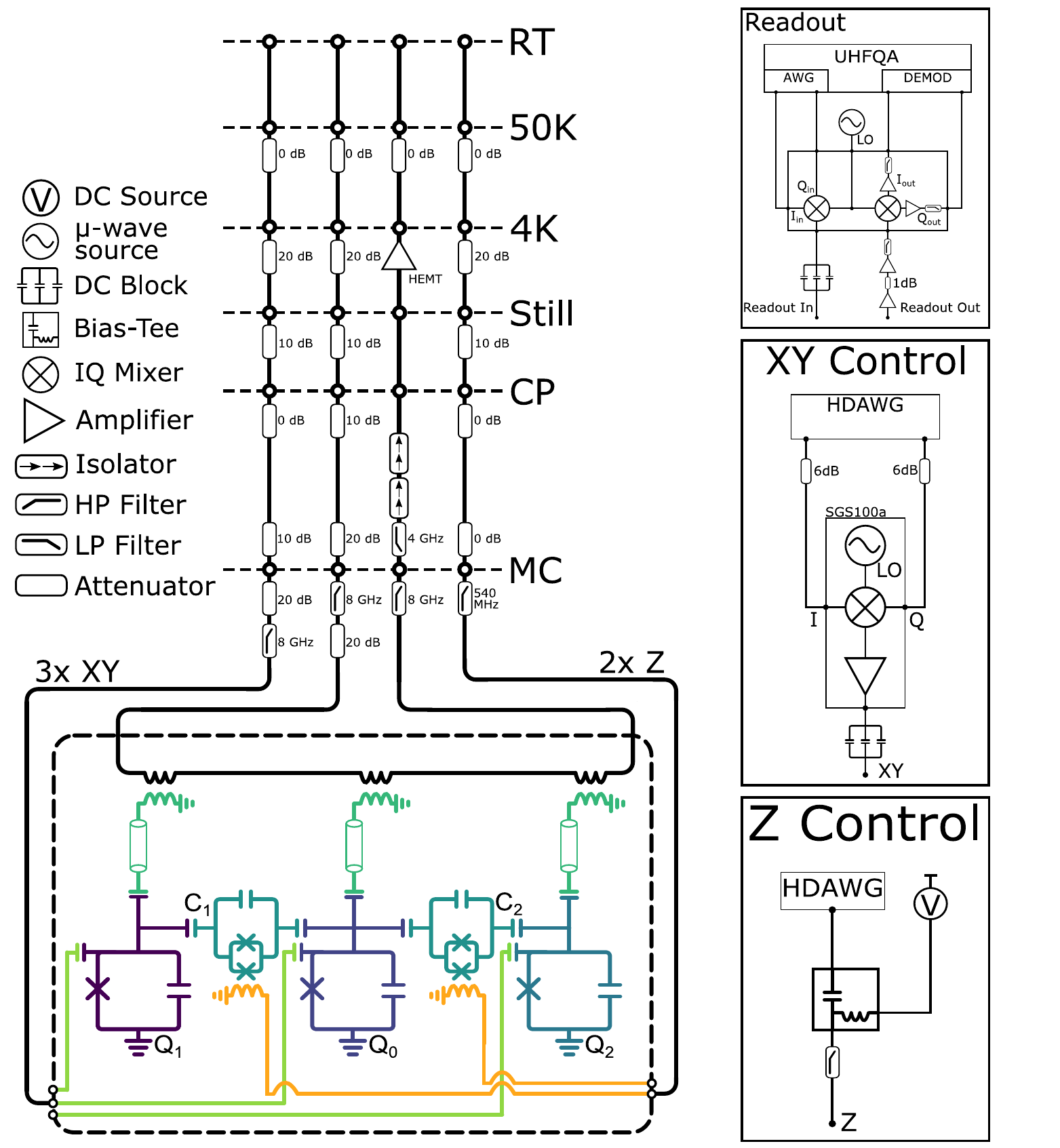}
    \caption{Cryogenic setup and full circuit of the three utilized qubits in the five qubit device. Compare the color coding to the  micrograph in \figpanel{fig:Fig1}{A}. Electronics setup for the room temperature control is to the right.}
    \label{fig:Setup}
\end{figure}
Qubit fabrication is performed as in previous work~\cite{Bengtsson2020}. Additionally, we make use of aluminum crossovers to aid in routing of signals across the device and for tying together ground planes of the chip. The device is packaged in a copper box and wirebonded to a palladium- and gold-plated printed circuit board (PCB). An aluminum shield with a volume cut out around the PCB traces and the chip is fixed atop the device to push package modes away from the operating frequencies of the device and provide an additional layer of shielding. The PCB contains 16 non-magnetic connectors of which we use seven: two for the input and output of the readout, three for local control of the single qubits, and two for the static and AC flux control of the couplers.

The setup used in this experiment is a standard circuit-QED setup. The copper package housing the sample sits at the bottom of our Bluefors LD250 dilution refrigerator and is shielded from magnetic fields by two shields of cryoperm/mu-metal and two superconducting shields. All signal lines are attenuated and filtered to thermalize the signals coming into the fridge.

We perform readout using a Zurich Instruments UHFQA for generating and reading out the signals. The readout pulses pass through an up/down-conversion board where the local oscillator (LO) from a Rohde \& Schwarz SGS100a continuous-wave signal generator is split between both the up- and down-conversion halves. This maintains the phase coherence between the generation and digitization of the readout signals. Single-qubit pulses are synthesized using a Zurich Instruments HDAWG and upconverted internally using Rohde \& Schwarz SGS100a vector signal generators. The flux drives for couplers is generated digitally by the HDAWG as the signal frequencies for our coupling gates fall within the bandwidth of the HDAWG. 

Importantly, we fix the trigger period of the measurements such that it is an even multiple of the least common multiple of the inverse of the LO frequencies of the qubits
\begin{equation}
    \tau_p = n \times \text{lcm}\left(\frac{1}{f_{LO}^0}, \frac{1}{f_{LO}^1}, \frac{1}{f_{LO}^2}\right)
\end{equation}
For qubit LOs of \unit{4.5}{GHz}, \unit{4.5}{GHz}, \unit{5}{GHz} this ends up being an even multiple of \unit{2}{ns}. In our case, we set our trigger period to \unit{350}{$\mu$s} to allow for adequate reset time. This ensures that every time we trigger a measurement the qubits see the same phase from shot to shot. All phase control of the pulses can then be handled by digitally manipulating the carrier of the pulses generated on the HDAWG. This holds for the adjustments in the local phases of the qubits to update the qubit frame with virtual-Z gates, and those of the flux drives on the couplers, allowing full control over the SWAP phase $\phi$ of the three-qubit gate.

\section{Qubit control and characterization}
We perform standard spectroscopy and coherence measurements for each qubit individually, to determine the readout frequencies, qubit/coupler frequencies, anharmonicities, decoherence rates, and coupling strengths.

After the initial characterization, we optimize the single-qubit pulses using first-order DRAG~\cite{PhysRevLett.103.110501} to produce \unit{20}{ns} high fidelity single-qubit pulses with a cosine envelope. We then optimize the single-shot readout fidelity for the $\ket{0}$, $\ket{1}$, and $\ket{2}$ states for each qubit, so that we can measure the population transfers of the two-qubit gate calibration and of the three-qubit gate. A linear discriminator is used to label and differentiate readout for the qubits.
\begin{table*}[ht]
    \begin{ruledtabular}
    %\color{red}
    \begin{tabular}{lccccc}
        Parameter           & Qubit 1    &Coupler 1 & Qubit 0 & Coupler 2 & Qubit 2 \\
        \hline
        $f_R$               & \unit{6.57}{GHz} & & \unit{6.89}{GHz} & & \unit{6.74}{GHz}\\
        $g$                 & \unit{57.0}{MHz} & & \unit{49.9}{MHz} & & \unit{59.8}{MHz}\\
        $f_{01}$            & \unit{4.18}{GHz} & & \unit{4.73}{GHz} & & \unit{4.30}{GHz}\\
        $f_{12}-f_{01}$     & \unit{-219}{MHz} & & \unit{-229}{MHz} & & \unit{-227}{MHz}\\
        $f_{C}|_{\Phi_b=0}$ &        & \unit{8.969}{GHz} & & \unit{8.69}{GHz} & \\
        $J$                 &           &55/\unit{54}{MHz}  & &55/\unit{55}{MHz}&  \\
        $T_{1}$             & \unit{27.35(384)}{$\mu$s}  & & \unit{39.73(719)}{$\mu$s} & & \unit{34.73(420)}{$\mu$s}\\
        $T_{2}^{*}$         & \unit{45.21(785)}{$\mu$s}  & & \unit{57.67(1319)}{$\mu$s}& & \unit{21.49(635)}{$\mu$s}\\
        \end{tabular}
        \end{ruledtabular}
        \caption{\label{tab:device_table}Device parameters. Readout-resonator frequency $f_R$ and qubit transition frequencies $f_{ij}$. $f_C$ is the estimated frequency of the coupler at zero flux bias. Qubit-resonator coupling $g$ and qubit-coupler coupling $J$. $T_1$ and $T_2^*$ are the relaxation and free induction decay times, respectively, measured over 12 hours at the operation point of the three-qubit gate.}
\end{table*}

We then find an optimal working point of the coupler bias for implementing the individual two-qubit gates. We first characterize the DC flux crosstalk between the two couplers finding a residual DC crosstalk of $<0.6\%$ of a flux quantum. 

We then apply two sequences to characterize the coupler/qubit spectrum. In the first, we generate states in the single-excitation manifold. We then apply a long ($\sim$\unit{2}{$\mu$s}) AC flux pulse to the coupler which is connected to an excited qubit. The frequency of the AC pulse is swept in a region around the target two-qubit gate frequencies as well as the DC bias of the coupler. This procedure finds all sidebands which could collide with the target interactions we want to drive. 

In the second sequence, we prepare either $\ket{101}$ or $\ket{110}$ to find the frequency with which to drive the individual CZ transitions comprising the three-qubit gate, as a function of the DC bias, on the respective couplers. This identifies a region in the bias landscape that is free from collisions between the single-excitation manifold and the two-excitation manifold as our operating points. Using these points, we then perform a recalibration of the qubit/resonator frequencies, single-qubit gates, the coherence times and other properties, see Table~\ref{tab:device_table}.
% \clearpage
\section{\texorpdfstring{$\Lambda$-system dynamics}{Lambda-system dynamics}}
Consider three states labelled $\psi_1, \psi_2, \psi_3$, where communication between states $\psi_1$ and $\psi_3$ is mediated through a common intermediate state $\psi_2$. This situation is described by \eqref{eqn:Eq2} (\figpanel{fig:Fig1}{C}), but we reiterate it here for clarity:
\begin{equation}
    \label{eqn:LambdaModel}
    H = \begin{bmatrix}
            -\delta_1 & g_1 & 0  \\
            g_1^* & \delta_1 - \delta_3 & g_3 \\
            0 & g_3^* & \delta_3
        \end{bmatrix},
\end{equation}
where $2\delta_i$ is the detuning between the shared transition $\psi_2$ and the corresponding state $i=\{1,3\}$. $g_i$ is the coupling strength between the participant eigenstates and the intermediate eigenstate. We introduce new parameters,
\begin{eqnarray}
    p~&&= |g_1|^2 + |g_3|^2 + \delta_1^2 + \delta_3^2 - \delta_1\delta_3, \\
    q~&&= |g_1|^2\delta_3 - |g_3|^2\delta_1 + \delta_1^2\delta_3 - \delta_3^2\delta_1.
\end{eqnarray}
This allows us to recast the characteristic polynomial of the Hamiltonian as a depressed cubic equation
\begin{equation}
    -x^3 + px + q = 0,
\end{equation}
where $p$ and $q$ are always real, even if the couplings are not. The solutions to the Schr\"odinger equation with this parameterization are
\begin{eqnarray}
   \psi_1(t) &&=\sum_{j=1}^{3} \frac{e^{ix_jt}}{3x_j^2-p}\Big(\psi_1^0(x_j^2+\delta_1 x_j - |g_3|^2 - \delta_3^2 + \delta_1\delta_3)-\psi_2^0 g_1(x_j+\delta_3)+\psi_0^3g_1g_3\Big),\\
   \psi_2(t) &&=\sum_{j=1}^{3} \frac{e^{ix_jt}}{3x_j^2-p}\Big(-\psi_1^0g_1^*(x_j+\delta_3) + \psi_2^0(x_j^2 + (\delta_3-\delta_1)x_j-\delta_1\delta_3)-\psi_3^0g_3(x_j-\delta_1)\Big),\\
   \psi_3(t) &&=\sum_{j=1}^{3} \frac{e^{ix_jt}}{3x_j^2-p}\Big(\psi_1^0g_1^*g_3^* - \psi_2^0(x_j-\delta_1) + \psi_3^0(x_j^2-\delta_3x_j-|g_1|^2-\delta_1^2 + \delta_1\delta_3)\Big),
\end{eqnarray}
where $x_1$, $x_2$, $x_3$ are the solutions of the characteristic polynomial with explicit form
\begin{equation}
    x_k = \frac{1}{3}\left(\xi^k F + \frac{3p}{\xi^k F}\right)
\end{equation}
and
\begin{eqnarray}
   \xi &&=\frac{-1+\sqrt{-3}}{2}, \\
   F   &&= \sqrt[3]{\frac{27q \pm \sqrt{27}\sqrt{27q^2-4p^3}}{2}}.
\end{eqnarray}
With these solutions we can compute the probability of being in state $\psi_k$ as a function of time as $P_k(t)=|\psi_k(t)|^2$. We use this formulation to calibrate and fit the parameters for the three-qubit gate of the main text (see \figpanel{fig:Fig1}{D}), where $\psi_1=\ket{110}$, $\psi_2=\ket{200}$, and $\psi_3=\ket{101}$.
\begin{figure}[ht]
    \centering
    \includegraphics[width=0.8\linewidth]{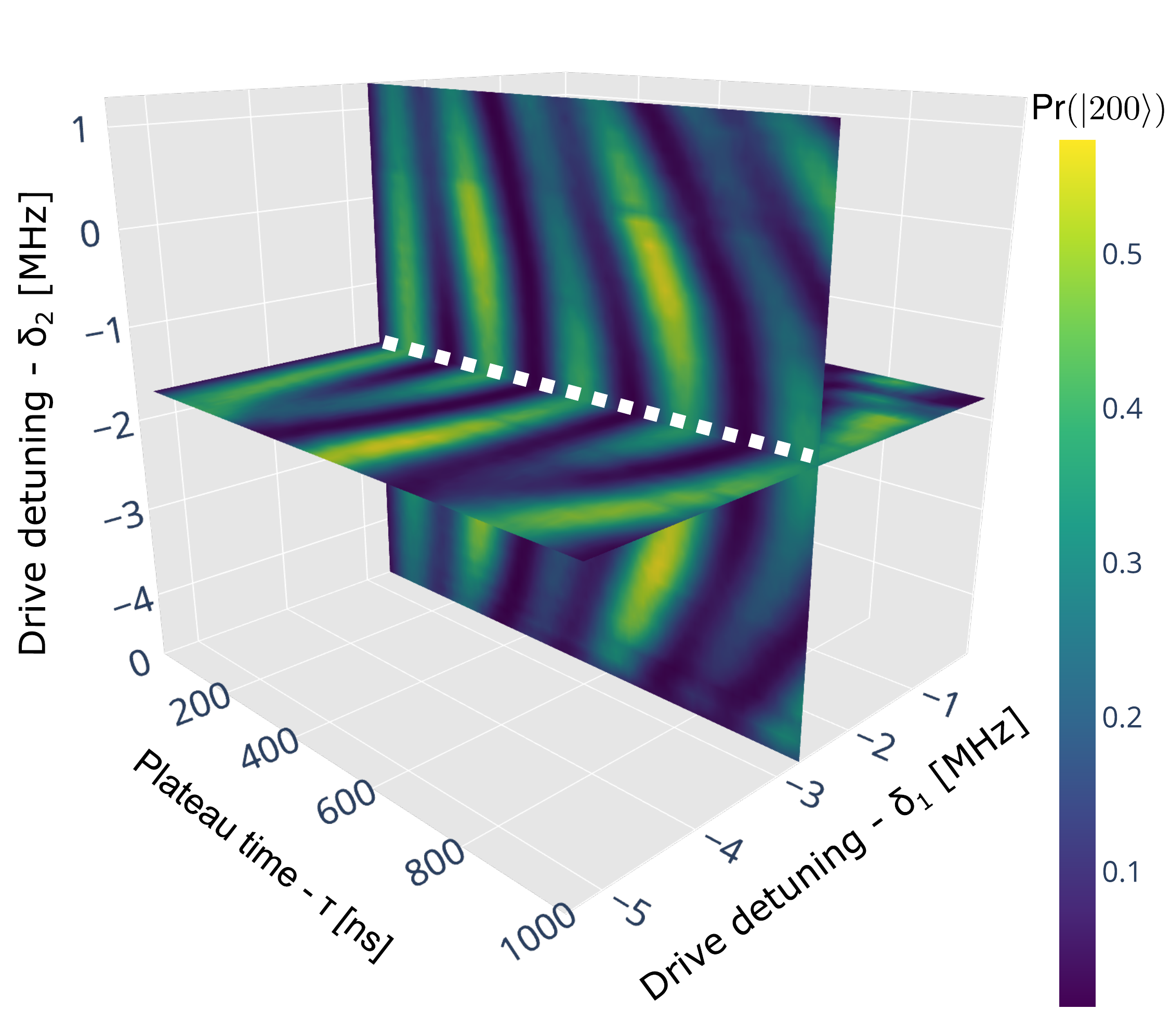}
    \caption{Experimental 3D cross sections of the three-qubit gate calibration. The plateau times of the simultaneous AC flux pulses are swept while fixing the frequency of the drive on one coupler and sweeping the frequency of the other. The roles are then reversed and the other coupler is swept. The intersection of the oscillations is where the ideal three-qubit gate is calibrated to $\theta=\pi/2$ and $\gamma=0$. The white dashed line corresponds to the linecut in  \figpanel{fig:Fig1}{D} presented in the main text.}
    \label{fig:LandScape}
\end{figure}
% \clearpage
\section{\texorpdfstring{Coherence limits of driven $\Lambda$-system}{Coherence limits of driven Lambda-system}}
Infidelity from decoherence can be computed, for low decoherence rates, by solving the master equation modeling the dissipative evolution of the system,
\begin{equation}
    \dot{\rho} (t) = - \frac{i}{\hbar} [H (t), \rho (t)] + \sum_{k} \gamma_k \mathcal{D} [L_k] \rho (t),
\end{equation}
where $\rho (t)$ is the system density matrix, $H (t)$ is the Hamiltonian, and $\mathcal{D} [L_k] \rho$ and $\gamma_k$ are, respectively, the Lindblad superoperator associated to the $k$th dissipation process and its coefficient, which encode the decoherence rates. The superoperator is
\begin{equation}
    \mathcal{D} [L_k] \rho = L_k \rho L_k^\dagger - \frac{1}{2} \left\{ L_k^\dagger L_k, \rho \right\}.
\end{equation}
In our case, we will only consider energy relaxation and dephasing.

The dynamics of the system can be approximately modeled with the following Hamiltonian, which extends the model of \eqref{eqn:LambdaModel} to include the three-excitation manifold:
\begin{align}
    H_{\rm eff} = \big[& \tilde{J}_{01} \left( \lvert 110 \rangle \langle 200 \rvert + \lvert 111 \rangle \langle 201 \rvert \right) \nonumber \\
    &+ \tilde{J}_{02} \left( \lvert 101 \rangle \langle 200 \rvert + \lvert 111 \rangle \langle 210 \rvert \right) + \mathrm{H.c.} \big] \nonumber \\
    &+ \delta \left( \lvert 200 \rangle \langle 200 \rvert - \lvert 111 \rangle \langle 111 \rvert \right) .
\end{align}
We have assumed that the oscillatory drives give rise to effective constant coupling strengths between states as in the main text and have neglected possible losses from leakage to the couplers. 

Focusing on the case under study, $U_{CCZS} (\pi / 2, \phi, 0)$, with total gate time equal to $\tau$, we define $\tilde{J}_{01} = \pi / \sqrt{2} \tau$, $\tilde{J}_{02} = - \tilde{J}_{01} e^{i \phi}$, and $\delta = 0$. Note that from the matrix structure of the Hamiltonian, the main mechanism of the gate consists of the two systems, $\{ \lvert 200 \rangle, \lvert 101 \rangle, \lvert 110 \rangle \}$ and $\{ \lvert 111 \rangle, \lvert 210 \rangle, \lvert 201 \rangle \}$, discussed elsewhere in this manuscript.

The three-qubit gate uses states outside of the computational subspace in its implementation, which forces us to model the qubits with at least three levels each. However, the choice of Lindblad jump operators and their rates becomes more involved when qutrits are considered instead of qubits ~\cite{korotkov2013error}. One could use the annihilation operator $a$ as jump operator, since it is the most natural generalization of $\sigma_-$, the usual jump operator for relaxation in qubits. However, we understand relaxation as two different processes described by $\sigma_-$-like matrices, one for $\ket{1} \to \ket{0}$ and another one for $\ket{2} \to \ket{1}$, with approximately twice the rate of the former due to the small anharmonicity. One could even add a third process for $\ket{2} \to \ket{0}$ with a different rate, but the single-photon transition rate for $\ket{2} \to \ket{0}$ is exponentially small in the transmon regime and the two-photon emission does not occur in the undriven system. The coefficients of the Lindblad master equation and jump operators associated to the different processes affecting each qubit are
\begin{alignat}{4}
    \gamma_{10} =&\ \Gamma_1 \, , & \,\,\,\, & L_{10} =&\ \left(
        \begin{array}{ccc}
            0 & 1 & 0 \\
            0 & 0 & 0 \\
            0 & 0 & 0
        \end{array}
    \right) ,  \\
    \gamma_{21} =&\ 2 \Gamma_1 \, , & \,\,\,\, & L_{21} =&\ \left(
        \begin{array}{ccc}
            0 & 0 & 0 \\
            0 & 0 & 1 \\
            0 & 0 & 0
        \end{array}
    \right) ,  \\
    \gamma_{\phi} =&\ \frac{\Gamma_\phi}{2} \, , & \,\,\,\, & L_{\phi} =&\ \left(
        \begin{array}{ccc}
            0 & 0 & 0 \\
            0 & 2 & 0 \\
            0 & 0 & 4
        \end{array}
    \right) ,
\end{alignat}
where we have expressed the coefficients in terms of the standard decoherence rates $\Gamma_k = 1 / T_k$, with $k = 1, \phi$. The coefficients can also be expressed in terms of the transverse relaxation rate $\Gamma_2$ according to the relation $\Gamma_2 = \Gamma_1 / 2 + \Gamma_\phi$. These rates and operators appear in the master equation separately for each qubit, and the three-qubit operators can be constructed from these ones by tensor products with identity operations on the other two qubits. As a consequence, we obtain the complete set of equations governing the time evolution of the system by substituting the appropriate decoherence rates and operators for each qubit in the sum. 

Solving the master equation can be done perturbatively if the decoherence rates are low~\cite{Abad2021}. The density matrix is expanded to linear order in the decoherence rates, $\rho (t) = \rho_0 (t) + \rho_1 (t) + \mathcal{O} \Big( (t\Gamma_k)^2 \Big)$, and an analogous expansion of the master equation gives the set of equations satisfied by the two density-matrix contributions
\begin{align}
    \label{eqn:Lindblad}
    \dot{\rho}_0 (t) =&\ - \frac{i}{\hbar} [ H_{\rm eff}, \rho_0 (t) ] \, , \\
    \label{eqn:Lindblad2}
    \dot{\rho}_1 (t) =&\ - \frac{i}{\hbar} [ H_{\rm eff}, \rho_1 (t) ] + \sum_k \gamma_k \mathcal{D} [ L_k ] \rho_0 (t). 
\end{align}
For the analysis, we restrict to the case where the system always starts in a pure initial state $\ket{\psi}$ with non-zero components only in the computational subspace. Then, the density-matrix terms satisfy initial conditions $\rho_0 (0) = \ketbra{\psi}{\psi}$ and $\rho_1 (0) = 0$.

With the solutions for the master equation, we compute the average gate fidelity~\cite{NIELSEN2002249}
\begin{equation}
    F_{av} = \frac{\sum_{j,k} \alpha_{jk} \mathrm{Tr} \left[ U U_j^\dagger U^\dagger \mathcal{E} (\rho_k) \right] + d^2}{d^2 (d + 1)}, 
\end{equation}
where $d$ is the dimension of the Hilbert space, $U$ is the ideal gate $U_{CCZS} (\pi, \phi, 0)$, $\mathcal{E} (\rho_k)$ is the density matrix that results from evolving the initial state $\rho_k$, and $U_j$ and $\alpha_{jk}$ are, respectively, a complete basis of operators and the matrix of change between this basis and the initial states $\rho_k$. The transformation is defined as $U_j = \sum_k \alpha_{jk} \rho_k$.

\eqrefs{eqn:Lindblad}{eqn:Lindblad2} can in principle be applied to processes acting on Hilbert spaces of any dimension $d$, but we only interpret processes as gates if they act in the computational subspace. To remedy this for the qutrit basis, we project the time-evolved density matrix at time equal to the total gate time $\tau$, $\mathcal{E} (\lvert \psi \rangle \langle \psi \rvert) \equiv \rho_0 (\tau) + \rho_1 (\tau)$, onto the computational subspace. 

Finally, to compute the average gate fidelity we select the complete basis to sum over. As we are working with a three-qubit Hilbert space, we substitute the labels $\{ j, k \}$ in (S19) by three-index labels $\{ ijk, mnp \}$. For initial states $\rho_{mnp} = \lvert \psi_{mnp} \rangle \langle \psi_{mnp} \rvert$, and using all the combinations of tensor products of the single-qubit basis,
\begin{equation}
    \psi_{mnp} = \lvert \psi_i \rangle \otimes \lvert \psi_j \rangle \otimes \lvert \psi_k \rangle \, , \, \mathrm{where} \, \lvert \psi_i \rangle \in \{ \lvert 0 \rangle , \lvert 1 \rangle , \lvert + \rangle , \lvert - \rangle \}.
\end{equation}
For the operator basis, we use the Pauli basis
\begin{equation}
    U_{ijk} = U_i \otimes U_j \otimes U_k \, , \, \mathrm{where} \, U_i \in \{ \mathbb{I}, \sigma_x, \sigma_y, \sigma_z \}.
\end{equation}
The choice of states and operators that preserve the tensor structure is a convenient one, because then the matrix of change of basis will also preserve the tensor structure,
\begin{equation}
    \alpha_{ijk,mnp} = \alpha_{im} \otimes \alpha_{jn} \otimes \alpha_{kp} \, , \, \mathrm{s.t.} \, U_j = \sum_k \alpha_{jk} \lvert \psi_k \rangle \langle \psi_k \rvert.
\end{equation}

\begin{figure}[ht]
    \centering
    \includegraphics[width=0.75\linewidth]{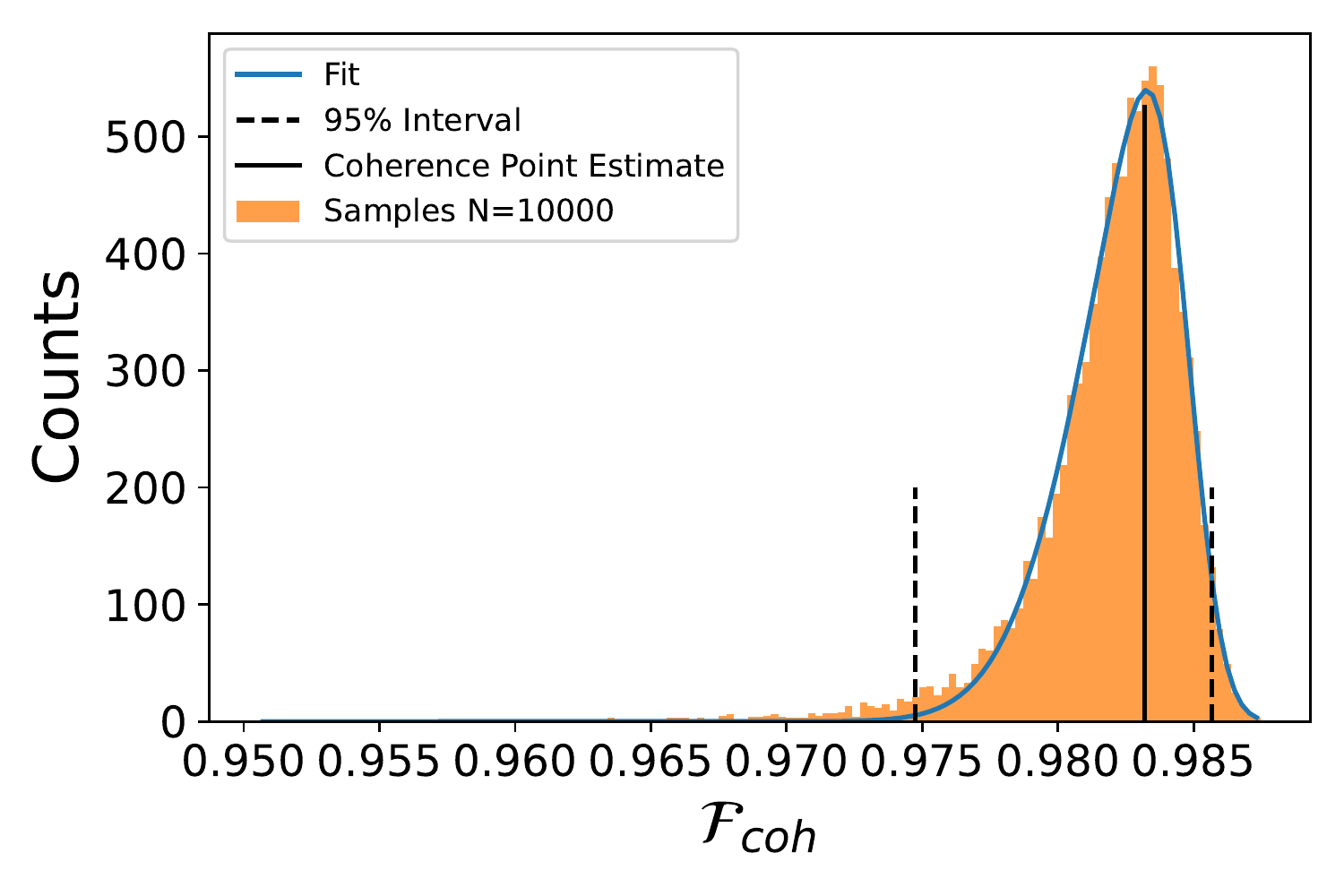}
    \caption{Monte-Carlo simulation of coherence limit of the parameters in Table~\ref{tab:device_table}. We report a $95\%$ interval from the resulting distribution.}
    \label{fig:S4}
\end{figure}

The final result for the average gate fidelity is
\begin{align}
    F_{av} = 1 &- \frac{5}{9} \Gamma_1^0 \tau - \frac{7}{18} \Gamma_1^1 \tau - \frac{7}{18} \Gamma_1^2 \tau \nonumber \\
    &- \frac{61}{72} \Gamma_\phi^0 \tau - \frac{125}{288} \Gamma_\phi^1 \tau - \frac{125}{288} \Gamma_\phi^2 \tau,
\end{align}
where the superscript on the decay rates denotes the qubit. This expression is transformed to the process fidelity $F_\chi$ by the use of the relation $1 - F_\chi = (1 - F_{av}) (d + 1) / d$. The resulting process fidelity is
\begin{align}
    F_\chi = 1 &- \frac{5}{8} \Gamma_1^0 \tau - \frac{7}{16} \Gamma_1^1 \tau - \frac{7}{16} \Gamma_1^2 \tau \nonumber \\
    &- \frac{61}{64} \Gamma_\phi^0 \tau - \frac{125}{256} \Gamma_\phi^1 \tau - \frac{125}{256} \Gamma_\phi^2 \tau .
\end{align}
It is this expression we use for computing the coherence limit for the three-qubit gate as shown in \figpanel{fig:Fig3}{D} and the parameters in Table~\ref{tab:device_table}. We use these values and perform a Monte-Carlo simulation of the coherence limit of the fidelity as shown in \figref{fig:S4}. We find that $95\%$ of the outcomes of the simulation lie in the range $[97.43\%,98.57\%]$ with a point estimate from the $T_1$ and $T_2$ values of $98.30\%$.
\section{Resource analysis of gate set tomography}
A natural question arises of why not just perform three-qubit gate set tomography and characterize everything. Here, we lay out a rough outline motivating the difficulty of GST experiment design and why we followed a method similar to~\cite{Geller2021}.

Given some work one can find 64 preparation fiducials, and 27 measurement fiducials corresponding to the same size of input as for process tomography. The number of preparation fiducials corresponding to an informationally complete set are set at $2^{2n}$ separate preparations. The number of measurement fiducials has a minimal number of circuits
\begin{align}
    N_{min}=\frac{2^{2n}-1}{2^n-1},
\end{align}
corresponding to the number of degrees of freedom for a trace-preserving matrix divided by the number of independent binary output strings that can result from a measurement. In our case, the lower bound number of measurement fiducials is 9. 

However, gate set tomography works by repeatedly amplifying errors generated by a set of operators known as germs~\cite{Nielsen2021gatesettomography}. This means that for every germ we would produce $64\times27=1728$ separate circuits. We were unable to perform germ selection to generate this set of amplificationally complete germ sequences for a static set of single-qubit gates and trace-preserving three-qubit gate due to memory constraints in the germ-selection algorithm. 

Despite these computational difficulties we can nonetheless estimate the resources necessary for an LGST sequence. For the three-qubit gate and the single qubit $R_x(\pi/2)$ and $R_y(\pi/2)$ gates the total number of circuits necessary for characterizing the operation would be $2^n \times 2^{2n} \times 3^n = 13824$ separate measurements for LGST. This is not an unreasonable number of circuits to run, but is nearly five times the number of circuits performed in our error mitigation technique. Going a step further and performing GST only on the bare three-qubit gate germ sequence up to a depth of $L=16$ would increase the total number of circuits to $>20$k and would only allow access to a restricted amount of information such as overrotations.

That is to say that without more exhaustive work on trying to trim down the number of circuits, we believe that our method is a satisfactory alternative to mitigating the errors of QPT. Process tomography can roughly be thought of as a single instance of an $L=1$ GST experiment over just one of the gates to be characterized in the gate set. What we have done is modify the GST protocol to separately characterize the single-qubit rotations and then use those to inform the $L=1$ three-qubit germ.

It would be interesting and fruitful to perform three-qubit GST  and find the optimal measurement fiducials, performing fiducial pair reduction, and looking at reduced models for the error generations~\cite{PRXQuantum.3.020335,Nielsen_2021}. However, three-qubit GST is beyond the scope of this work where we are simply trying to mitigate SPAM from the reconstruction and will be the subject of follow-up work.

\section{Single-qubit gate set tomography}
\begin{figure}
    \centering
    \includegraphics[width=0.75\linewidth]{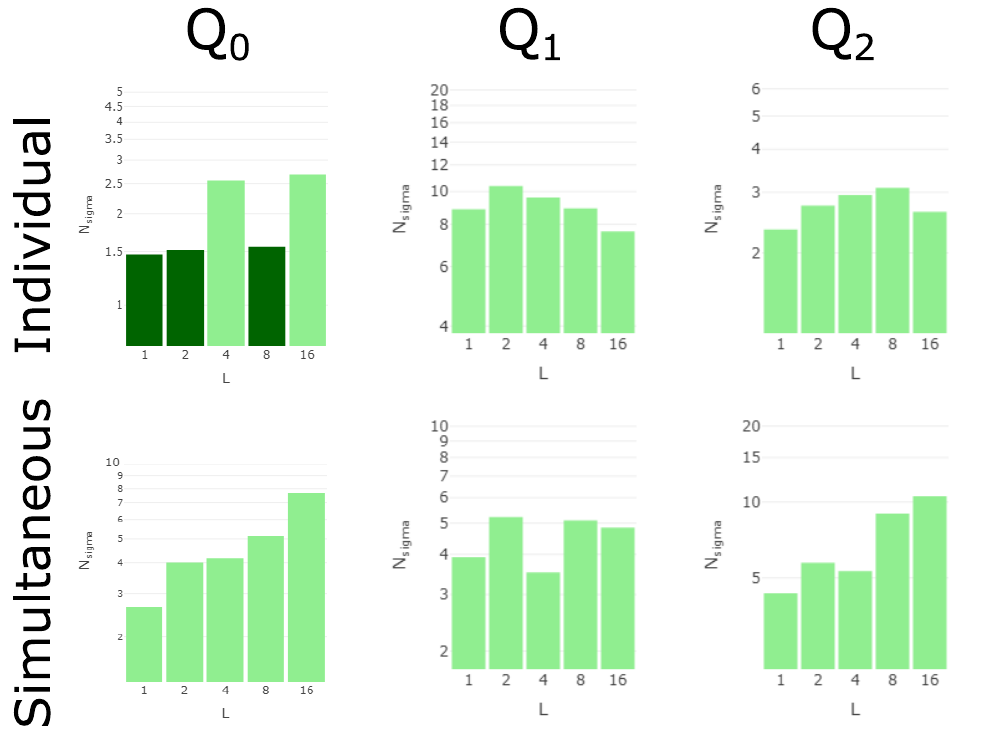}
    \caption{Model violations for running the long-sequence gate set tomography first individually with one of the qubits performing GST while the others idle and simultaneously performing GST across all qubits.}
    \label{fig:model_viol}
\end{figure}
For the single-qubit gate set tomography, we first set about verifying the independence of single-qubit operations and measurements. We perform GST individually across all qubits and then perform single-qubit GST simultaneously. In doing so we seek to find whether there is significant non-Markovianity between the two modes of operation. While we do find greater model violations between individual and simultaneous GST (\figref{fig:model_viol}) the reconstructed operations and infidelities are similar between the two runs. Non-Markovianity can also arise from drifts in parameters between the experiments as well as fluctuations in the coherence of the device which will occur over the runtime of the GST measurements. 

We perform a long-sequence GST (LSGST) set consisting of gates from the set $\{I, R_y(\pi), R_y(\pm \pi/2), R_x(\pm \pi/2)\}$. The total number of circuits performed in the analysis was 2904 separate gate sequences up to a depth of $L=16$ and each circuit was sampled $n=5000$ times. The circuits were generated and results were analyzed using pyGSTi~\cite{pygsti}. From this we extract for each qubit: the noisy initial states, $\tilde{\rho}_0$, noisy rotation operators, $\{\tilde{R}_x(\pm\pi/2), \tilde{R}_y(\pm\pi/2), \tilde{R}_y(\pi)\}$, and POVMs ${\tilde{M}_j}$ for mitigating the SPAM errors in the reconstruction. The results are summarized in Table~\ref{tab:gst_errors} and~\ref{tab:gate_errors}.
\begin{table}[ht]
    \begin{ruledtabular}
    \renewcommand{\arraystretch}{1.25}
    \begin{tabular}{l | c c c}
               & \large$Q_0$ & \large$Q_1$ & \large$Q_2$  \\
        \hline
        &&&\\
        \large$\tilde{\rho}_0$   
            & $\begin{bmatrix}  0.98267 & -0.0011495 \\ 0.0007066 &  0.01733 \end{bmatrix}$   
            & $\begin{bmatrix}  0.9786449 & -0.0001291 \\ 0.0008762 &  0.0213551 \end{bmatrix}$   
            & $\begin{bmatrix}  0.9831357 & -0.0025322 \\ -0.0009186 &  0.0168643 \end{bmatrix}$     \\
            &&&    \\
        \large$\tilde{M}_{\ket{0}}$   
            & $\begin{bmatrix}  0.9979474 & -0.0010535 \\ 0.0012261 &  0.0293532 \end{bmatrix}$    
            & $\begin{bmatrix}  0.9982515 & -0.0001658 \\ 0.0023416 &  0.0273934 \end{bmatrix}$   
            & $\begin{bmatrix}  0.9959302 & -0.0021059 \\ -0.0013668 &  0.0967759 \end{bmatrix}$     
    \end{tabular}
    \end{ruledtabular}
    \caption{Results from the simultaneous GST for the noisy initial states $\tilde{\rho}_0$ and $\tilde{M}_{\ket{0}}$.}
    \label{tab:gst_errors}
\end{table}

\begin{table}[ht]
    \begin{ruledtabular}
    \renewcommand{\arraystretch}{1.25}
    \tiny
    \begin{tabular}{l | c c c}
               & \large$Q_0$ & \large$Q_1$ & \large$Q_2$  \\
        \hline
        &&&\\
        \large
        $\tilde{R}_y(\pi)$   
            & $\begin{bmatrix}  1.0 & 0.0& 0.0 & 0.0\\
                                -0.000217 & -0.998879 & -0.024224 & -0.0008303\\
                                -0.000006 & -0.024439 & 0.999304 & 0.000972\\
                                -0.000062 & 0.000932 & 0.000294 & -0.999224\\
              \end{bmatrix}$   
            & $\begin{bmatrix}  1.0 & 0.0 & 0.0 & 0.0\\
                                -0.000010 & -0.999098 & -0.023026 & -0.000654\\
                                0.000096 & -0.024074 & 0.998841 & -0.000609\\
                                -0.000120 & -0.000082 & -0.000013 & -0.998913\\
              \end{bmatrix}$    
            & $\begin{bmatrix}  1.0 & 0.0 & 0.0 & 0.0\\
                                -0.000209& -0.999618 & -0.007606 & 0.002197\\
                                0.000118 & -0.007273 & 0.999349 & 0.004348\\
                                0.000712  & -0.002601 & 0.004141 & -0.999163\\
              \end{bmatrix}$    \\
            &&&    \\
        \large
        $\tilde{R}_y(\frac{\pi}{2})$   
            & $\begin{bmatrix}  1.0 & 0.0 & 0.0 & 0.0\\
                                0.000383 & -0.031803 & 0.012059 & 0.998644\\
                                0.000165 & 0.010784 & 0.999220 & -0.011690\\
                                0.000620 & -0.998535 & 0.010439 & -0.031815\\
              \end{bmatrix}$   
            & $\begin{bmatrix}  1.0 & 0.0 & 0.0 & 0.0\\
                                0.000755 & -0.042896 & 0.012600 & 0.997547\\
                                -0.000963 & 0.012871 & 0.997011 & -0.018112\\
                                0.001227 & -0.996440 & 0.011780 & -0.045188\\
              \end{bmatrix}$    
            & $\begin{bmatrix}  1.0 & 0.0 & 0.0 & 0.0\\
                                0.000901 & -0.022064 & 0.008149 & 0.998821\\
                                0.000338 & 0.000063 & 0.999143 & -0.008496\\
                                -0.000696 & -0.998992 & -0.000270 & -0.021328\\
              \end{bmatrix}$    \\
            &&&    \\
        \large
        $\tilde{R}_x(-\frac{\pi}{2})$ &
              $\begin{bmatrix}  1.0 & 0.0 & 0.0 & 0.0\\
                                0.000158 & 0.999369 & -0.012767 & 0.012619\\
                                -0.000136 & -0.013556 & -0.029575 & 0.998334\\
                                0.000321 & -0.013413 & -0.998543 & -0.029368\\
              \end{bmatrix}$   
            & $\begin{bmatrix}  1.0 & 0.0 & 0.0 & 0.0\\
                                0.000002 & 0.999659 & -0.011865 & 0.010912\\
                                -0.000052 & -0.01216 & -0.044557 & 0.998164\\
                                -0.000271 & -0.011201 & -0.997972 & -0.044769\\
              \end{bmatrix}$    
            & $\begin{bmatrix}  1.0 & 0.0 & 0.0 & 0.0\\
                                -0.000307 & 0.999171 & -0.003745 & 0.003989\\
                                0.000061 & -0.002733 & -0.018287 & 0.997997\\
                                0.000412 & -0.002978 & -0.997771 & -0.017622\\
              \end{bmatrix}$    \\
    \end{tabular}
    \end{ruledtabular}
    \caption{Results from GST for the noisy rotation operators in the Pauli-Product superoperator representation. For $\tilde{R}_y(-\pi/2)$ and $\tilde{R}_x(\pi/2)$ we assume that the opposite sign rotation is idealized by taking the Hermitian conjugate of the superoperator.}
    \label{tab:gate_errors}
\end{table}

\section{SPAM-independent process tomography}
In the process tomography~\cite{Chuang1997}, we first prepare the input probe states $\{\ket{g}, \ket{e}, \ket{+}, \ket{i^+} \}^{\otimes 3}$ by performing single-qubit rotations from the set $\{I, R_y(\pi), R_y(\pi/2), R_x(-\pi/2)\}^{\otimes 3}$. This gives us a set of 64 input states on which we apply the process to be characterized. Finally, we rotate the outcome into the bases $\{X,Y,Z\}^{\otimes 3}$ by applying single-qubit rotations from the set $\{R_y(-\pi/2), R_x(\pi/2), I\}^{\otimes 3}$. This choice of rotations maintains the parity of the eigenvalue associated with each input probe state when measured in its basis so that we are left with binary output strings in the space $\{0, 1\}^{\otimes 3}$, simplifying the model of the POVMs.

Using the results of GST we redefine the probe states in terms of the noisy initial states for the three qubits, $\tilde{\rho}_0=\bigotimes_{k=0}^2\tilde{\rho}_0^k$. We then prepare the input probe states with the noisy rotation operators, $\tilde{\textbf{R}}_i =\bigotimes_{k=0}^2 \tilde{R}^k$, to generate each of the 64 input states,
\begin{equation}
    \tilde{\rho}_i = \tilde{\textbf{R}}_i \tilde{\rho}_0 \tilde{\textbf{R}}_i^\dagger.
\end{equation}
The process is then applied to the state using the Choi matrix according to \eqref{eqn:eq8}. The resulting state is then rotated into its measurement basis and projected onto the set of outcomes $\{0, 1\}^{\otimes 3}$. We can represent our rotated POVM for a measurement outcome $s\in[1,8]$ and a particular Pauli basis $j\in[1,27]$ as
\begin{equation}
    \tilde{M}_{js} = \tilde{R}_j \tilde{M}_s \tilde{R}_j^\dagger.
\end{equation}
The probability that a three-qubit state has an outcome $s$, given a measurement basis $j$, and preparation $i$ is then
\begin{align}
    p_{i,j,s} = Tr(\tilde{M}_{j,s}\tilde{\rho}_i ') &= Tr(\tilde{M}_{j,s}\otimes I_d (\tilde{\rho}_i\otimes I_d)\rho_\Phi) \nonumber\\
    &= Tr((\tilde{M}_{j,s}\tilde{\rho}_i\otimes I_d)\rho_\Phi).
\end{align}
The probabilities can be written down as a vector and, since the above equation is linear, can be set up as a linear inversion problem that obtains the Choi matrix by inverting the equation
\begin{equation}
    A \vec{\rho_\Phi} = \vec{p}.
\end{equation}
Here, the matrix $A$ contains all the information regarding the probes and measurements with $\vec{\rho_\Phi}$ and $\vec{p}$ as the flattened Choi matrix and the probabilities. The construction of $A$ is described in Ref.~\cite{Knee2018}; we use a simple linear inversion of A to obtain and initial estimate of the process.
\section{Error analysis of reconstruction}
We perform bootstrapping on the experimental data to determine confidence intervals for our process and state reconstruction~\cite{Efron1993,Home2009}. This involves taking the empirically observed probability distributions and resampling from them the same number of times as was performed in the experiment. These newly sampled datasets represent possible observed outcomes at the level of the sampling error, assuming that our original dataset represents the underlying true distribution. For both the quantum process tomography and the state tomography, we take the resampled datasets and feed them into their respective reconstruction techniques, as previously outlined, and repeat the sampling a number of times, $N_{boot}=10^3$. The distribution of these reconstructions are then used to report the error on the obtained fidelity.

\begin{figure}[t]
    \centering
    \includegraphics[width=\linewidth]{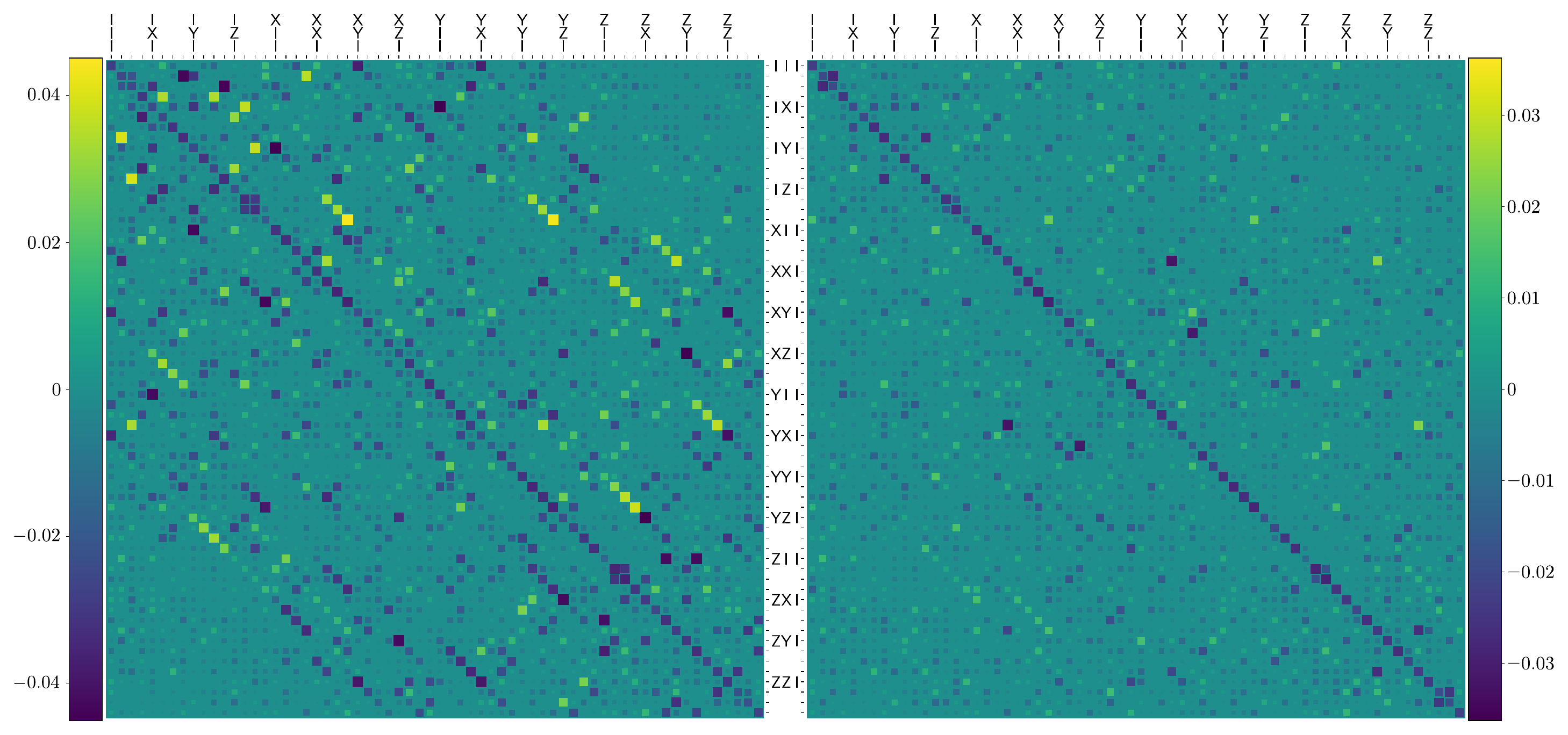}
    \caption{Plotting $\mathcal{E}_{ideal}^\dagger\mathcal{E}_{recon}-I$ for the ideal unitary $U_{CCZS}(\pi/2, \pi/2, 0)$, and $\tilde{\mathcal{E}}_{ideal}^\dagger\mathcal{E}_{recon}-I$ for the error mitigated unitary $U_{CCZS}(\tilde{\theta}, \tilde{\phi}, \tilde{\gamma})$ that most closely matches the reconstruction. $\mathcal{E}$ is the Pauli transfer matrix of the quantum process for the three-qubit gate.}
    \label{fig:PTM_comparison}
\end{figure}

In addition to the bootstrapping for the process tomography, to find the control-error-free fidelity, we reconstruct the process and perform an optimization to find the angles $(\tilde{\theta}, \tilde{\phi},\tilde{\gamma})$ of an ideal CCZS gate which best matches the reconstruction (see Table \ref{tab:fid_table} of the main text). We report this fidelity as being independent of calibration errors to isolate between control errors, coherence errors, leakage errors, and errors due to parasitic terms in the Hamiltonian. To test the validity of this assumption we compare the overlaps of the Pauli transfer matrices (PTM), \figref{fig:PTM_comparison},
\begin{equation}
    \mathcal{D} = \mathcal{E}^{\dagger}_{ideal}\mathcal{E}-I.
\end{equation}
The relative sparsity of the overlap for the control-error-free unitary, $U_{CCZS}(\tilde{\theta},\tilde{\phi},\tilde{\gamma})$, versus that of the ideal target unitary hints that much of the error can be ascribed to these miscalibrations. After these we find that single-qubit Pauli Z errors make up the next largest elements of the matrix.
% \clearpage
% \section{Sensitivity of gate parameters}
\section{Quantum State Reconstruction}

For the state tomography we prepare the states as the insets in \figpanel{fig:Fig4}{B} and \figpanelNoPrefix{fig:Fig4}{D} show and similarly measure the $\{X,Y,Z\}^{\otimes 3}$ basis. We apply the measurement mitigation we obtain from the measured POVMs and perform a maximum-likelihood estimation (MLE) to reconstruct the density matrix constraining the MLE to be trace-preserving. The density matrix is represented using the Cholesky decomposition making the reconstruction manifestly positive semidefinite,
\begin{equation}
    \rho = \frac{T^\dagger T}{\tr{T^\dagger T}}.
\end{equation}
T is a triangular matrix,
\begin{equation}
    T= \begin{bmatrix}
            t_0                            & 0                      & 0  & \dotsi & 0 \\
            t_{2^n}+it_{2^n+1}             & t_1                    & 0  & \dotsi & 0 \\
            t_{3(2^n-1)+1}+it_{3(2^n-1)+2} & t_{2^n+2} + it_{2^n+3} & t_2& \dotsi & 0 \\
            \vdots                         & \vdots                & \vdots & \ddots & 0\\
            \dotsi                         & \dotsi                 & \dotsi & t_{3(2^n-1)-1}+it_{3(2^n-1)} & t_{2^n-1}
       \end{bmatrix},
\end{equation}
where $\textbf{t}=[t_0, t_1, \dotsi, t_{4^n-1}]$ is a set of parameters containing all real numbers $t_i$.

For the state tomography, we perform post-processing on the relative phases between populations of the $\ket{\text{GHZ}}= 1/\sqrt{2}(\ket{000}+e^{i\phi}\ket{111})$ and $\ket{\text{W}}=1/\sqrt{3}(\ket{100}+e^{i\phi_1}\ket{010}+e^{i\phi_2}\ket{001})$ by finding local rotations $R_Z(\phi_i)$ for each reconstruction which maximize the overlap with their ideal states. Neither of these local operations alter the entanglement characteristics of the resulting states.

\begin{figure*}[ht]
    \includegraphics[width=0.9\linewidth]{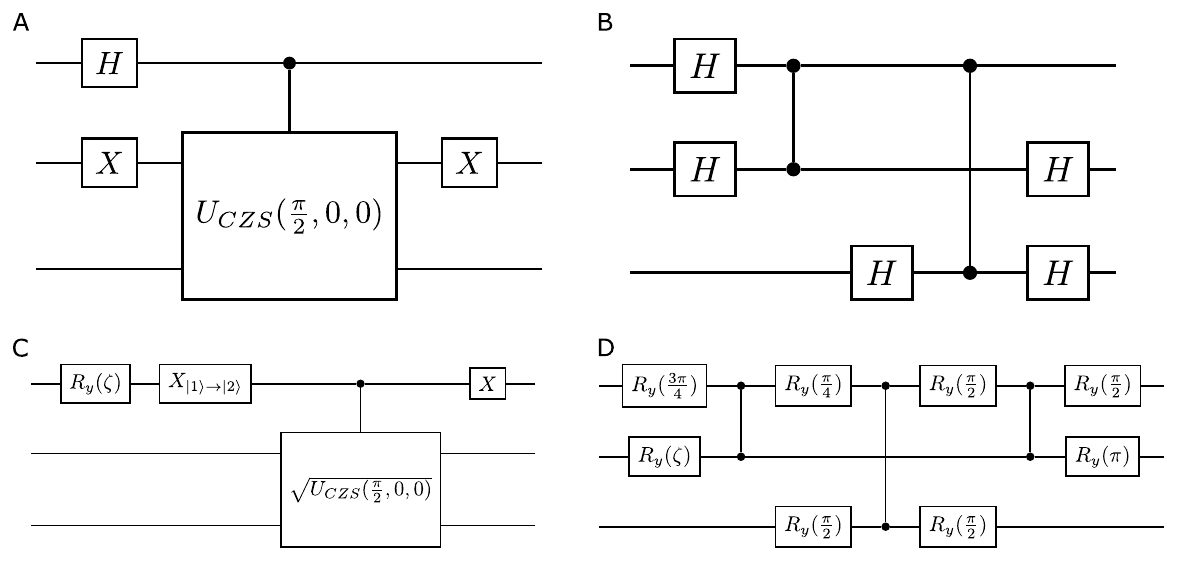}
    \caption{\textbf{A} and \textbf{C} GHZ- and W-state preparation circuits using the three-qubit CCZS gate. \textbf{B} and \textbf{D} Optimal compilation of the same circuits to two-qubit CZ gates. Preparation angle for the W-state, $\zeta=2\arccos\sqrt{1/3}$.}
    \label{fig:circuitcomp}
\end{figure*}

\figref{fig:circuitcomp} shows the circuits for generating the GHZ and W states from using either the native three-qubit CCZS gate or by using an optimal decomposition to two-qubit CZ gates. When predominantly coherence-limited, the three-qubit gate allows for substantial shortening of the circuit depth and compilation strategies.
% \clearpage

% \nocite{*}

\end{document}